\documentclass[12pt]{article}
\input epsf.sty

\def\lromn#1{\uppercase\expandafter{\romannumeral#1}}

\usepackage{graphicx,amsmath,amssymb}
\begin{document}

\begin{center}
\begin{large}
\textbf{
Light Propagation
and Paired Superradiance
in Coherent Medium
}

\end{large}
\end{center}

\vspace{2cm}
\begin{center}
\begin{large}
M. Yoshimura

Center of Quantum Universe, 
Faculty of Science, Okayama University, \\
Tsushima-naka 3-1-1, Kita-ku, Okayama,
700-8530 Japan

\end{large}
\end{center}

\vspace{3.5cm}
\begin{center}
\begin{Large}
{\bf ABSTRACT}
\end{Large}
\end{center}

The problem of light propagation of frequency corresponding to
half of the energy difference between
a metastable excited state and the ground state
of atoms is examined, and solved  for coherent medium
by analytic means.
We demonstrate that 
the non-linear system of Maxwell-Bloch equation for the effective model 
of the $\Lambda-$type three levels
is integrable in the mathematical sense.
Analytic solutions thus obtained describe pulse splitting accompanied by
compression, indicating a kind of non-linear
instability of propagating pulses.
The instability is eventually terminated by 
coherent two photon emission (called paired superradiance  
or PSR in short).
These results are displayed by
numerical outputs for visual understanding, as well. 
It is further shown that the integrable system allows 
a new class of soliton solutions.
Solitons, implying the phenomenon of seff-induced transparancy 
at non-resonant frequencies, are stable against PSR.
One of our goals of the present work is  construction of
a calculable theoretical framework for PSR rates associated 
with a trigger pulse propagation, which is achieved by
combining analytic results with perturbative methods.
PSR photon spectrum and its rate $\propto$(target number density)$^2$, 
along with their time structure,
are clarified this way.
These results may open a new path for
interesting technological applications such as
quantum entanglement and for solving
the remaining problems of the still mysterious neutrino.
Some basic strategy for realistic experiments of PSR detection and
soliton production is also outlined.

\newpage

%%%%%%%%%%%%%%%%%%%%%%%%%%%%%%%%%%%%%%%%%%%%%%%%%%%

{\bf \lromn1 \; Introduction}

\begin{figure*}[htbp]
 \begin{center}
 \epsfxsize=0.4\textwidth
 \centerline{\epsfbox{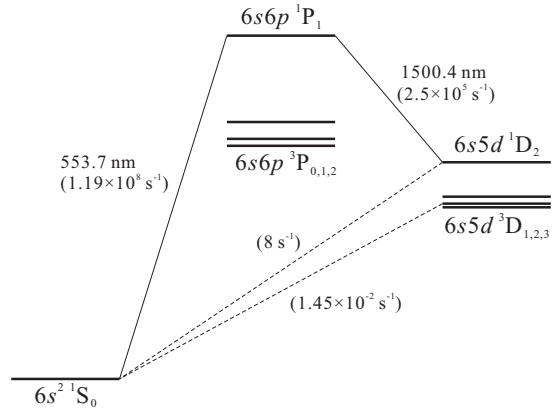}} \hspace*{\fill}
   \caption{$\Lambda-$type low lying levels of neutral Ba atom.
   An excitation scheme to $^1 D_2$ is also illustrated.
}
   \label{ba level}
 \end{center} 
\end{figure*}

Cooperative phenomenon known as superradiance (SR)
\cite{sr review}
may have more dramatic effects when applied to the
forbidden transition:
the decay rate via two photon emission 
may be enhanced to a macroscopic level 
without the wavelength restriction \cite{macro-coherence}.
This is in contrast to the single photon SR enhanced decay, with
the coherence region limited by
the wavelength$^2$.
We hereafter call  the macro-coherent two photon
emission \cite{macro-coherence} as paired superradiance (PSR).
An example of candidate atoms for PSR is
the first excited D-state of two electron system of Ba (its levels shown in
Fig(\ref{ba level})).

In the course of establishing a firm
theoretical formulation for PSR, 
we came accross a basic reference \cite{narducci}
(and quite possibly many related papers unknown to the present
author) in which the Maxwell-Bloch or rather a
Maxwell-Schr\"{o}dinger equation for an effective two
level problem is derived
and some numerical simulation based on this equation
is performed.

The purpose of the present paper is four-fold;
(1) to add to the literature of this field,
fully integrated analytic solutions of the relevant 
non-linear system, which describe pulse splitting
and compression, (2) to present a new class of soliton
solutions of quantized pulse area (the terminology
to be defined) that describe the phenomenon 
of self-induced transparancy(SIT),
(3) to give the PSR rate and  its spectrum, both time
dependent, along with its relation to soliton formation,
(4) to provide the signal to the noise ratio (S/N) for
radiative neutrino pair emission RNPE
(a new, proposed experimental method for the precision neutrino
mass spectroscopy, \cite{my-06}, \cite{pv}).

Our analytic solutions describe the multiple pulse splitting
accompanied by compression of propagating pulses
in a long target medium, when the input pulse is strong enough.
The pulse strength is made quantitative by our own concept of the pulse
area, a product of integrated field flux and the coupling strength to the medium.
The phenomenon is a highly non-linear coherent effect
of the whole system of target atoms and fields.

The multiple pulse splitting in a long target
may further be interpreted as a process towards formation of many solitons,
if the effect of PSR is included.
Indivisual split pulses in medium become increasingly sharper
prior to PSR.
The ever sharpening pulse has an ever increasing
energy density of fields and is unstable against 
the physical process of PSR.
It must become stabilized by PSR emission, eventually resulting
in formation of solitons, objects stable against PSR.
This intuitive picture has been supported by extensive numerical
computations performed by the present author, some of which
are shown below.
We would like to convince even the uneducated reader of this simple
picture of what occurs in an ideal envirornment,
the infinitely long coherent medium.

Solitons of the quantized pulse area and
of arbitrary velocities have much simpler analytic forms than in
the case of two level system \cite{coherent light propagation in 2 level}.
We give explicit formulas and exhibit their pulse shape in a number of figures.
Moreover,
SIT occurs at non-resonant frequencies, hence
might be more useful in technological applications
such as communication by light and quantum entanglement: 
excellent candidates  are (J=0 $\rightarrow$ J=0 ) PSR transitions
in alkhali earth atoms.

Our general method also allows to discuss residual interaction
between two solitons,
which turns out attractive, suggesting existence of bound states of
two solitons, presumably even a possibility of field condensate.
Implication of the attractive nature of force shall be addressed
in forthcoming work.

We next compute, based on perturbative methods, PSR rates 
in which one of the photons is the
forward going pulse component.
In PSR, two photons are, almost exactly, back to back emitted and
have the same energy, or
only this configuration in the two photon phase space
is macroscopically enhanced.
Calculated rates are more than adequate for
detection, and even give a hope of PSR
measurement in ion traps where one can  expect to store
a total target number of ions only as large as $10^6$.

The time structure of our triggered PSR is complicated.
PSR  is expected to occur rapidly and violently, 
most dramatically just prior to the soliton formation
(and at the time of its artificial destruction).
After soliton formation PSR stops, and 
propagating pulses become stabilized as solitons.
An ideal, and our own favorite, method of observing PSR is
creation of as many as possible solitons and their subsequent
controlled destruction, giving the largest PSR rates
at the instants of creation and destruction.
We would also like to stress, for PSR detection and
soliton production, great advantages of non-resonant frequency
of the trigger laser, which is non-destructive to target atoms
and makes experiments easier.

In any of these phenomena, irradiation
of the triggering laser at the doubled wavelength
is of vital importance to us, and we must fully understand
the problem of light propagation in medium,
which is done below.
In the literature \cite{sr review}
that deals with SR of the two level atom,
SR initiated by quantum fluctuation 
has extensively been discussed.
The initial setting  that interests us most in the present work
is considerably different from this initial
condition. PSR initiated by quantum fluctuation is weaker and
moreover it is our main intention to utilize benefits of the triggering laser.
Thus, our problem is more akin to the triggered or
induced SR in the two level problem \cite{sr review}.
Related to this, the time delay observed in fluctuation initiated
SR is absent in our triggered PSR.
Despite of all these, we shall briefly mention
interesting features of almost trigger-less PSR under a weaker trigger; 
excellent experimental signatures along with a rough estimate of
its rate.

In most of our analysis below, we ignore
relaxation processes, due to that the enhanced
PSR is very fast:  this seems
a legitimate approximation under a wide range of
circumstances.
Moreover, the coherence we need for PSR 
is not a stationary state of target atoms, but
rather it is a dynamically developed (with time) state
of the whole system of target atoms plus fields due
to the non-linear interaction between the two system.
In actual experiments,
the coherence is dynamically generated during a short time interval
via a series of physical
processes of excitation and trigger by lasers.
If the time of coherence development is
shorter than a typical relaxation time of relevant target state,
one could achieve a well prepared state for PSR.
In this sense what is needed for a positive measurement
is, in addition to the fast PSR rate, a fast
preparation of coherence development,
whose realization is left to a challenge
for experimentalists.

At a fundamental level the soliton formation may give rise
to a controlled measurement of RNPE, \cite{my-06}, \cite{pv},
because solitons are stable against two photon
emission, regarded as a crucial background to RNPE.
Solitons are, on the other hand,  
unstable against RNPE, hence soliton formation enhances
the signal to the background ratio in favor of RNPE.
If the enhanced RNPE rate is larger than (spontaneous
decay) rates of
the next leading order QED processes (usually 
much smaller than 1 msec$^{-1}$ order),
then the PSR background suppressed RNPE (due to
two effects discussed in the present work; 
mismatched trigger frequency to PSR and the soliton
formation) is measurable by well controlled experiments.

The present paper is organized as
a collection of (hopefully) compact sections
containing many numerical outputs,
and two long mathematical appendices:
section headings are 
\lromn2 Effective two level model,
\lromn3 Maxwell-Bloch equation and equation for
the tipping angle of Bloch vector,
\lromn4 Construction of analytic solutions,
\lromn5 Soliton solutions,
\lromn6 Theory of PSR and its relation to soliton formation,
%\lromn7 Numerical outputs for illustration of physics behind,
\lromn7 Outlook for RNPE,
\lromn8 Appendix \lromn1 Derivation of effective two level model,
\lromn9 Appendix \lromn2 Details towards construction of analytic solutions.
We present  numerical outputs, presumably more than
necessary to the educated reader, to help even the uneducated reader 
easily understand physics behind analytic and numerical results.
Appendices give some lengthy details of derivation omitted in the main
body of the text.

It is hoped that this work helps experimentalists to
design clever methods of detecting and measuring PSR and of
creating  optical soliton of our type.

Throughout this work we use the natural unit so that
$\hbar =1\,, c=1$.

%%%%%%%%%%%%%%%%%%%%%%%%%%%%%%%%%%%%%%%%%%%%%%%%%%%

\vspace{1cm}

{\bf \lromn2 \;
Effective two level model}

In this section
we follow \cite{narducci}
to sketch derivation of the coupled non-linear partial differential
equations of Maxwell-Bloch (MB) type, with a slight
modification of notation and correction of mistakes
at detuned frequencies.
The essence is to derive an effective two level model for
atoms of $\Lambda-$type levels such as Fig(\ref{ba level}).

The spirit of this model is a long time average,
or a truncation of past memory effects, 
in the sense of Markovian approximation, and
slowly varying envelope approximation (SVEA).
These should be an excellent approximation in our light propagation
problem in medium, because
this light wave at the doubled wavelength is non-resonant
to target atoms, its energy being 
too far away from the frequency to excite to the first atomic level above the ground.

The model is summarized by the Schr\"{o}dinger equation
for two amplitudes of two lower levels; a S-state, $^1S_0$ (its amplitude
given by $c_g$) and a  D-state, $^1D_2$ (given by $c_e$) in
the case of Ba atom.
It has the structure of two $\times$ two quantum
system of an effective Hamiltonian ${\cal H}$;
\begin{eqnarray}
&&
\frac{d}{dt} \left(
\begin{array}{c}
c_{e}  \\
c_{g}  
\end{array}
\right)= - i{\cal H}\left(
\begin{array}{c}
c_{e}  \\
c_{g}  
\end{array}
\right)
\,,
\\ &&
{\cal H} =
\left(
\begin{array}{cc}
 \mu_{ee}|E_0|^2 & e^{-i(2\omega - E_{eg})t  } \mu_{eg} (E_0^*)^2  \\
e^{i(2\omega - E_{eg})t  } \mu_{ge} E_0^2 &   \mu_{gg}|E_0|^2
\end{array}
\right)
\,.
\label{effective hamiltonian}
\end{eqnarray}
Here $E_0(x,t)$ is the slowly varying envelope of fast oscillating
field (to be multiplied by $e^{i\omega t}$).
We only discuss one dimensional propagation problem, assuming
the axial symmetry around the propagation axis,
taken here as $x-$axis.

Coupling strength here is given in terms of electric 
(or magnetic) dipole moments between the upper level $|j \rangle$; 
P-states in the Ba case
(the leading candidate being $^1P_1$), denoted by
$d_{je}\,, d_{jg}$;
\begin{eqnarray}
&&
 \mu_{ee} = 2 \sum_j \frac{d_{je}^2E_{je}}{E_{je}^2 - \omega^2}
\,, \hspace{0.5cm}
 \mu_{gg} = 2 \sum_j \frac{d_{jg}^2E_{jg}}{E_{jg}^2 - \omega^2}
 \,,
\\ &&
 \mu_{eg} = \sum_j \frac{d_{je}d_{jg} }{E_c - \delta \omega}
\,, \hspace{0.5cm}
 \mu_{ge} = \sum_j \frac{d_{je}d_{jg} }{E_c + \delta \omega}
\\ &&
E_c = \frac{1}{2}(E_{jg} + E_{je})
\,, \hspace{0.5cm}
\delta \omega = \omega - \frac{E_{eg}}{2} 
\,.
\end{eqnarray}
E1 or M1 moments $d_{ji}$ are related to experimentally measured
(or theoretically calculated in some unfortunate cases)
decay rates $\gamma_{ji}$ by $d_{ji}^2 = 3\pi \gamma_{ji} /E_{ji}^3$ 
with $E_{ji}$ the known energy difference between two levels.
Higher levels denoted by $|j \rangle$  are assumed
connected to two lowest levels, $|e \rangle$ and $|g \rangle$, 
by strong transition elements.
It would often be sufficient to consider a single level for $|j\rangle$
of strongest coupling.
As explained in Appendix \lromn1,
one may assume  to a good approximation at $\omega \sim E_{eg}/2$ the symmetry 
$ \mu_{eg} =  \mu_{ge} $, which we take in the rest of the text.

Matrix elements in ${\cal H}$,
eq.(\ref{effective hamiltonian}), 
of our effective $\Lambda-$model are due to the Stark effect.
It is important to recognize that
this shift contains two powers of propagating field $E_0^2$,
its absolute value being the power or the flux of propagating field.
The magnitude of Stark shift is not negligible for strong
laser irradiation: two diagonal elements are of order, $6$GHz for the D-state and 
$16$ GHz for the S-state, and 2.1 GHz
for the magnitude of off-diagonal elements,
assuming the laser power of $10^6$W mm$^{-2}$.

Derivation and neglected terms of this Hamiltonian structure are discussed in
our Appendix \lromn1.

\vspace{1cm}

{\bf \lromn3 \; Maxwell-Bloch equation and equation for
the tipping angle of Bloch vector 
}

The next step is to derive a macroscopic coupled set of
equations for polarization of medium and propagating field.
Throughout this work
we take the continuum limit of distributed
target atoms, and assume an axial symmetry around
the axis of pulse propagation, which
reduces our problem  to $1+1$ (one time $t$ and one space coordinate $x$)
dimensional field theory.
Polarization of medium is defined by
a Bloch vector of 3 components
(a part of density matrix elements $R_{\alpha \beta}$
in the notation of Appendix \lromn1);
\begin{eqnarray}
&&
R_1 = i n (c_g^*c_e e^{i\eta } - c_e^* c_g e^{-i\eta })
\,,
\\ &&
R_2 = - n (c_g^*c_e e^{i\eta } + c_e^* c_g e^{-i\eta })
\,,
\\ &&
R_3 = n (|c_e|^2 - |c_g|^2)
\,,
\\ &&
\eta = (2\omega -E_{eg}) t - 2kx + 2\varphi
\,.
\end{eqnarray}
Here $n(x)$ is the number density of targets
assumed at rest, and
we took for simplicity a constant density $n(x) = n$ within a medium
of finite length, $0\leq x \leq L$.
The function $\omega t - kx + \varphi(x,t)$
is the main phase part of the
propagating field $e^{i\omega t}E_0$.

From the Schr\"{o}dinger equation we derive
the Bloch equation for polarization components,
\begin{eqnarray}
&&
\frac{\partial}{\partial t}R_1 
=  \frac{ \mu_{ee} -  \mu_{gg}}{4}|E_0^2| R_2 + \frac{\mu_{ge}}{2}|E_0^2| R_3 
\,,
\\ &&
\frac{\partial}{\partial t}R_2 = -  \frac{ \mu_{ee} -  \mu_{gg}}{4}|E_0^2| R_1
\,,
\\ &&
\frac{\partial}{\partial t}R_3 = - \frac{\mu_{ge}}{2} |E_0^2| R_1
\,.
\end{eqnarray}
From this set of equations one has a conservation of the magnitude
of the Bloch vector,
\begin{eqnarray}
&&
\frac{\partial}{\partial t} ( R_1^2 + R_2^2 + R_3^2) = 0
\,,
\end{eqnarray}
hence $\vec{R}^2(x,t)$ is time independent, which can be taken the
squared number density $n^2(x)$.

A linear combination of $R_2$ and $R_3$,
\begin{eqnarray}
&&
R_2'  = \frac{R_2 - \gamma R_3}{\sqrt{1 + \gamma^2}} 
\,,
\hspace{0.5cm}
\gamma = \frac{ \mu_{ee} -  \mu_{gg}}{2\mu_{ge}}
\,,
\end{eqnarray}
is  conserved in our system of differential equations, 
and one may set this vanishing without any
loss of generality, giving the condition $R_2 = \gamma R_3$.
The Bloch equation is thus effectively reduced to
\begin{eqnarray}
&&
\frac{\partial}{\partial t}R_1 
=  \frac{ \mu_{ge}}{2}(1+\gamma^2)|E_0^2| R_3 
\,,
\label{bloch 1}
\\ &&
\frac{\partial}{\partial t}R_3 = - \frac{\mu_{ge}}{2} |E_0^2| R_1
\,,
\label{bloch 2}
\end{eqnarray}
with a constant of motion,
$R_1^2 + (1+\gamma^2)R_3^2 =n^2(x)$.
For the Ba D-state, $|\gamma| \sim 2.3$.

Strictly, the Bloch vector has 4 components, but
to a good approximation the other component is time independent.
We shall ignore this complication, relegating some
explanation to Appendix \lromn1.

\begin{figure*}[htbp]
 \begin{center}
 \epsfxsize=0.6\textwidth
 \centerline{\epsfbox{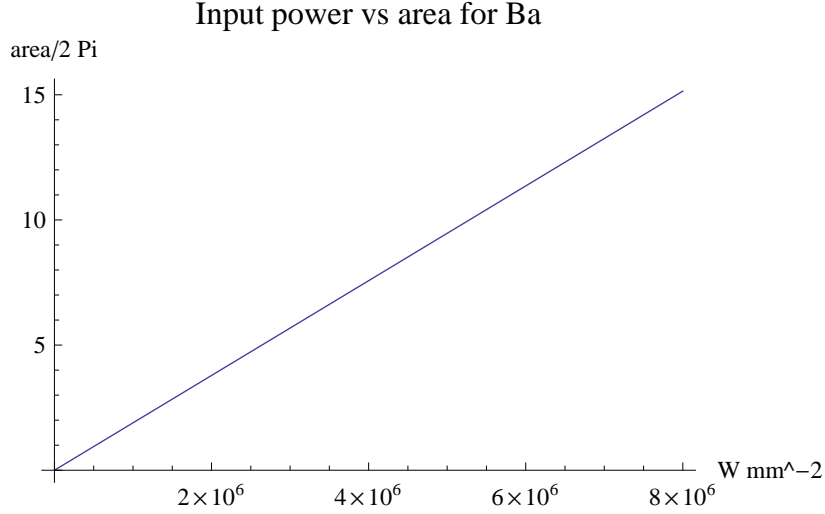}} \hspace*{\fill}
   \caption{Pulse area of Ba D-state vs the input pulse power.
   Quantized areas of an integer number $\times 2\pi$
   can give rise to this number of solitons.
   Pulse duration 3 ns assumed.
}
   \label{Area vs input laser intensity}
 \end{center} 
\end{figure*}

To proceed further, it is convenient to introduce the 
angle function $\theta(x,t)$, called the tipping angle, and the constant of motion 
$B = \pm n(x)$ by
\begin{eqnarray}
&&
R_1 = B \sin \theta
\,, \hspace{0.5cm}
R_3 = \frac{B}{\sqrt{1+\gamma^2}}  \cos \theta
\,.
\label{tipping angle def}
\end{eqnarray}
The Bloch equation is then equivalent to a relation
between $\theta$ and the field power $|E_0^2|$;
\begin{eqnarray}
&&
\partial_t \theta = \omega_R \,,
\hspace{0.5cm}
\omega_R \equiv 
\tilde{\mu} |E_0^2| 
\label{area relation}
\,,
\\ &&
\tilde{\mu} \equiv \frac{\sqrt{1 + \gamma^2}} {2} \mu_{ge} 
= \frac{1}{4}\sqrt{(\mu_{gg}-\mu_{ee})^2 + 4 \mu_{ge}^2}
\,,
\label{tilde mu def}
\end{eqnarray}
with $\tilde{\mu} |E_0^2| \sim 3 {\rm GHz} |E_0^2|/(10^6 {\rm W mm}^{-2})$ 
for the Ba D-state.
The tipping angle $\theta$ is thus an integrated flux of
$|E_0^2|$, with
a weight given by the strength of Stark shift $\tilde{\mu}$,
\begin{eqnarray}
&&
 \theta(x,t) = \tilde{\mu}\int_{-\infty}^{t-x} dy 
|E_0^2(x, y)|
\,.
\label{def tipping angle}
\end{eqnarray}
$\theta(x,\infty)$ is called the pulse area,
while $\theta(x,t)$ the area function in the present work.
The relation (\ref{def tipping angle}), 
containing the squared amplitude $|E_0^2|$, is different 
from the corresponding one in the two level problem
\cite{sr review}
in which the field $E_0$ itself appears in the integrand.
As an illustration,
the area $\theta(x=3 {\rm cm}\,, t=\infty)$ is plotted 
as a function of input laser power
in Fig(\ref{Area vs input laser intensity}),
which we need to find out necessary power values
for subsequent computations on the Ba D-state.
Moreover, the pulse area of $2\pi$ is an
important unit for quantized solitons,
a subject fully discussed in the following sections.

It is important to distinguish two cases of different signs of the constant $B$.
If $B = n> 0$, the  population difference $R_3$ is positive
for small $\theta$, which means that there  are more atoms in
the excited state than in the ground state.
If $B = -n < 0$, there  are more atoms in
the ground state than in the excited state.
From obvious reasons, we use the terminology similar 
to the case of the two level problem:
the terminology of
amplifier is used for $B>0$ and the absorber for $B<0$.
Needless to say, this distinction is interchanged by
a redefining transformation $\theta \rightarrow \theta + \pi$.
Nevertheless, it is customary to take the limit value 
of the tipping angle,
$\theta(-\infty) \rightarrow 0$, hence we
stick to this terminology.

Polarization $R_i$ of medium is related to the field $E_0$
by the Maxwell equation.
We discuss right-moving field given by
$|E_0(x,t)| e^{-i \omega (t -x) }$.
Under SVEA, the evolution
of envelope $|E_0(x,t)|$ is given by 
\begin{eqnarray}
&&
(\partial_t + \partial_x)|E_0|^2 = \omega \mu_{ge} |E_0^2| R_1
\,.
\label{maxwell}
\end{eqnarray}
The coupled system of non-linear partial differential equations,
eq.(\ref{bloch 1}), eq.(\ref{bloch 2}) and eq.(\ref{maxwell}),
forms the fundamental equation of field
propagation in medium, which may be called the Maxwell-Bloch equation
for our effective $\Lambda-$model.

The Maxwell-Bloch equation becomes an equation for
the  single variable $\theta(x,t)$, using eq.(\ref{tipping angle def});
\begin{eqnarray}
&&
(\partial_t + \partial_x)\partial_t \theta = \pm \alpha \sin \theta
\partial_t \theta
\,,
\hspace{1cm}
\alpha \equiv \omega \mu_{ge} n
\,.
\label{basic eq}
\end{eqnarray}
Our basic problem is then to solve this single, non-linear equation
under an arbitrary initial and boundary data.
Difference from the two level problem is in the RHS term;
in the two level case there is no $\partial_t \theta$ and only term 
$\propto \sin \theta$, which is a potential term
in the ordinary sense.
This is the familiar sine-Gordon equation for propagation of pulses
and SR at the resonant frequency \cite{sr review}.

A simplest form of relaxation may be introduced 
into the envelope equation (\ref{maxwell})
as a friction of the form, $\propto \kappa |E_0^2|$.
This leads to a modified equation for $\theta$;
\begin{eqnarray}
&&
(\partial_t + \partial_x)\partial_t \theta 
+ \kappa \partial_t \theta = \pm \alpha  \sin \theta
\partial_t \theta
\,.
\label{basic eq 2}
\end{eqnarray}
Both forms of equation, eq.(\ref{basic eq})
and eq.(\ref{basic eq 2}), may be integrated once,
introducing an arbitrary function $A(x)$ of space coordinate;
\begin{eqnarray}
&&
(\partial_t + \partial_x) \theta 
+ \kappa  \theta  \pm \alpha  \cos \theta
= A(x)
\,.
\label{basic eq once integrated 0}
\end{eqnarray}

There are more complicated forms of relaxation such as
the inhomogeneous Doppler broadening in gas \cite{doppler broadening},
which has to be treated separately.

When the initial data obeys the condition,
$|A(x)| > \alpha $,
the envelope may grow unlimitedly.
This is a situation we do not discuss
as our physics problem.
With $|A(x)| < \alpha $, we set $A(x) = \pm \alpha \cos \theta_i (x)$. 
For the dissipationless case of $\kappa = 0$,
the once-integrated equation becomes
\begin{eqnarray}
&&
(\partial_t + \partial_x) \theta 
\pm \alpha  (\cos \theta - \cos \theta_i)
= 0
\,.
\label{basic eq once integrated}
\end{eqnarray}

\vspace{1cm}

{\bf \lromn4 \; Construction of analytic solutions
}

It would be instructive first to
discuss homogeneous solutions without
spatial $x$ dependence.

\begin{figure*}[htbp]
 \begin{center}
 \epsfxsize=0.6\textwidth
 \centerline{\epsfbox{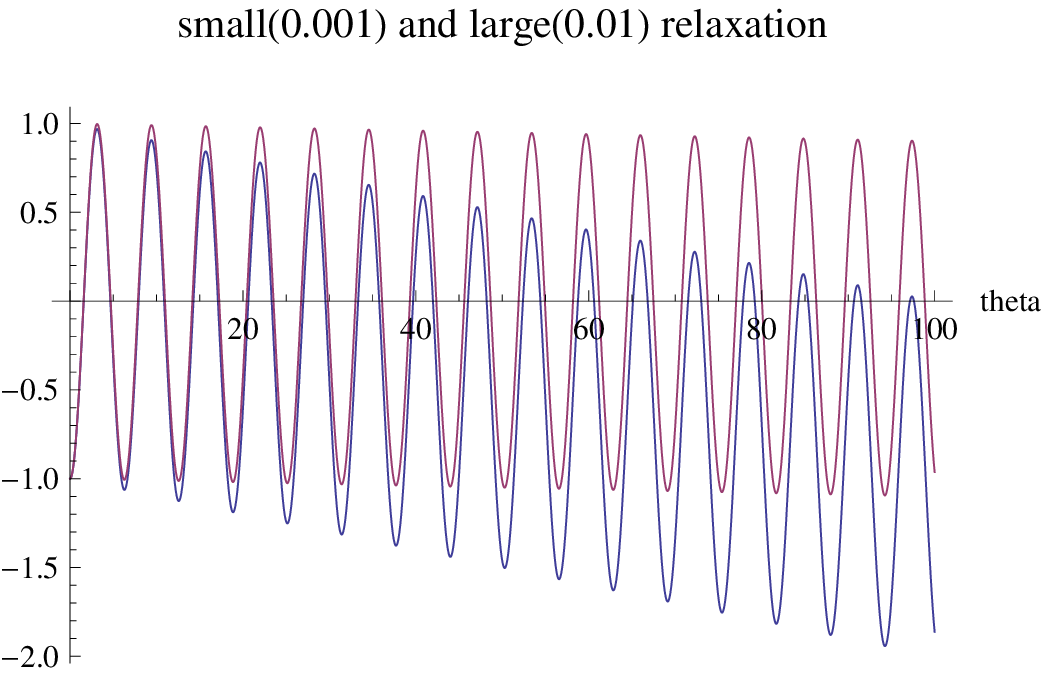}} \hspace*{\fill}
   \caption{Allowed islands of the tipping angle
given by positive ordinate values.
   Two choices of friction, $\kappa/\alpha = 10^{-2}$ (in blue)
and $10^{-3}$  (in purple), are compared.
Smaller frictions give more islands, giving a formation
chance of more solitons.
}
   \label{theta velocity}
 \end{center} 
\end{figure*}

The ordinary differential equation
with the simplest relaxation included is
given by
\begin{eqnarray}
&&
\frac{d^2 \theta }{dt^2} +(\kappa \mp \alpha \sin \theta ) \frac{d \theta }{dt} = 0
\,.
\label{theta eq no space}
\end{eqnarray}
This describes a dynamical system of a fictitious pendulum (its angle
location given by $\theta$)
with the friction term
varying periodically with $\theta$, $\kappa \mp \alpha \sin \theta$.
There is no force acting in the ordinary sense.
Equation (\ref{theta eq no space}) may be once integrated to give the
velocity $d\theta/dt = -(\kappa \theta \pm \alpha \cos \theta)$, 
after a suitable redefinition of $\theta$ variable.
The function of $\theta$, 
$-(\kappa \theta/\alpha +  \cos \theta)$, is plotted in Fig(\ref{theta velocity})
for large and small frictions,
to illustrate importance of the magnitude $\kappa /\alpha$.
The region of $d\theta/dt\geq 0$ alone is
allowed by the energy positivity $|E_0^2| \geq 0$,
and this restricts the $\theta $ region to be
given by an inequality, $-(\kappa \theta  + \alpha \cos \theta) \geq 0$,
forming islands of $\theta$ values nearly periodically separated 
for small $\kappa /\alpha$.
For the dissipationless case of $\kappa = 0$,
there are infinitely many pieces of such $\theta-$islands,
while these regions are limited in number for
a finite $\kappa$.
This dynamically allowed number of $\theta-$islands gives
the possible maximal number of produced solitons,
as clarified later.

The magnitude of $\alpha$ for Ba is $\sim 0.72$ GHz$\,n/(10^{16}$cm$^{-3})$.
Smaller values of friction, e.g., less than $10^{-2}\alpha$
(the factor $10^{-2}$ somewhat arbitrarily chosen), imply
relaxation times larger than $\sim 140$ ns $(n/10^{16}$cm$^{-3})^{-1}$.
We may also compare these values with the Doppler broadening
in Ba gas, $\Delta_D \sim 680$MHz$(T/300 K)^{1/2}$ (with $T$
the temperature).
The requirement $\Delta_D < \alpha$ gives $T < 340 K (n/10^{16}{\rm cm}^{-3})^2$.
We assume below that $\kappa\,, \Delta_D$ and all other
relaxation rates (including those in solids) are
much smaller than $\alpha$, and consider the dissipasionless case.

The explicit solution without space coordinate dependence is, 
in the dissipationless case of $\kappa=0$,
\begin{eqnarray}
&&
\cos \theta = - \frac{\cos \theta_0 \cosh \left(
\alpha (t-t_0)\sin \theta_0\right) - 1}{\cosh \left( \alpha (t-t_0)\sin \theta_0\right) - \cos \theta_0}
\,,
\label{area-no-space}
\\ &&
\omega_R =
\frac{d\theta }{dt} = \frac{ \alpha \sin^2 \theta_0 }{\cosh \left( \alpha (t-t_0) \sin \theta_0\right)- \cos \theta_0}
\,,
\\ &&
R_3 = - n \frac{\cos \theta_0 \cosh \left(
\alpha (t-t_0)\sin \theta_0\right) - 1}{\cosh \left( \alpha (t-t_0)\sin \theta_0\right) - \cos \theta_0}
\,.
\end{eqnarray}
This solution describes a dynamic motion of $\theta$ that
starts from $\theta_0$ at $t = -\infty$,
reaches $\pi$ at $t=t_0$, and ends finally at $2\pi - \theta_0 $.

The method of deriving more general solutions of the full partial differential
equation is to let parameters $t_0\,, \theta_0$ 
here to depend on other variables of $t-x\,, x$.
Details of this construction are given in 
Appendix \lromn2.

Simplest solutions that describe the initial target state either
fully in the ground state (denoted by $^{(g)}$ and called the absorber) 
or in the excited state 
(denoted by $^{(e)}$ and called the amplifier) are
given by
\begin{eqnarray}
&&
|E_0^2(x,t)|^{(e)}  
= \frac{\epsilon_0^2(t-x)}{1 - \alpha x \sin \tilde{\theta}
 + (\alpha x)^2\sin^2 (\tilde{\theta}/2)}
 \,, 
\label{amplifier: ground 0}
\\ &&
R_3^{(e)} \sim n
\frac{\cos \tilde{\theta} - \alpha x \sin\tilde{\theta}
+ (\alpha x)^2 \sin^2 (\tilde{\theta}/2) }
{1 - \alpha x \sin \tilde{\theta}
 + (\alpha x)^2\sin^2 (\tilde{\theta}/2)}
 \,,
\\ &&
|E_0^2(x,t)|^{(g)}  
= \frac{\epsilon_0^2(t-x)}{1 + \alpha x \sin \tilde{\theta}
 + (\alpha x)^2\sin^2 (\tilde{\theta}/2)}
 \,,
\\ &&
R_3^{(g)} \sim - n
\frac{\cos \tilde{\theta} + \alpha x \sin\tilde{\theta}
+ (\alpha x)^2 \sin^2 (\tilde{\theta}/2) }
{1 + \alpha x \sin \tilde{\theta}
 + (\alpha x)^2\sin^2 (\tilde{\theta}/2)}
 \,,
\label{absorber population}
\\ &&
\tilde{\theta}(t-x) = \tilde{\mu} \int_{-\infty}^{t-x} dy \epsilon_0^2(y)
\,.
\label{area function}
\end{eqnarray}
$\epsilon_0^2(t-x)$ here is the input pulse intensity.

\begin{figure*}[htbp]
 \begin{center}
 \epsfxsize=0.6\textwidth
 \centerline{\epsfbox{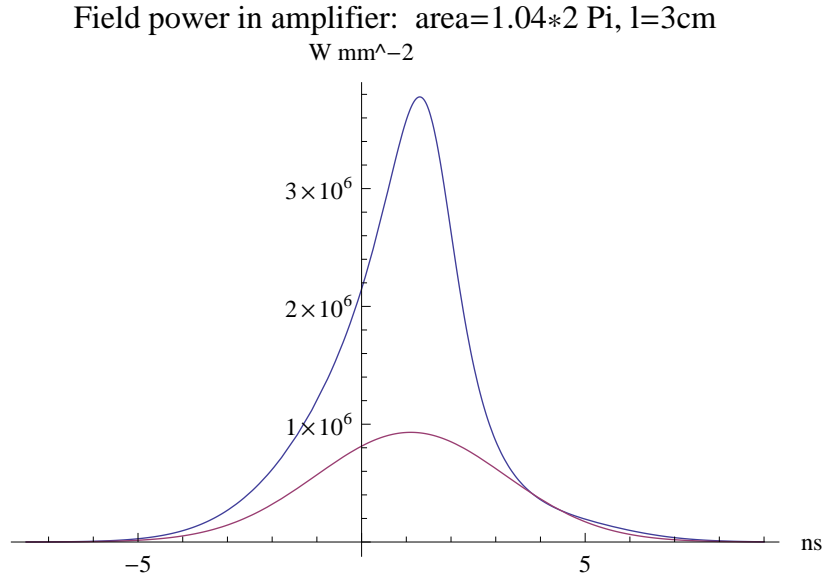}} \hspace*{\fill}
   \caption{Compressed pulse (in blue) compared with incident pulse (in purple).
   The time delay (measured in difference of central locations of the two peaks) 
common in SR is absent here.
   The absolute value of abscissa coordinate is meaningless.
Number density $10^{17}$cm$^{-3}$, target length = 3 cm, laser
   of power $\sim 2\pi$ pulse,
   duration 3 ns, and the initial angle $\theta_0=\pi/4$ are assumed. 
}
   \label{Pulse compression}
 \end{center} 
\end{figure*}

\begin{figure*}[htbp]
 \begin{center}
 \epsfxsize=0.6\textwidth
 \centerline{\epsfbox{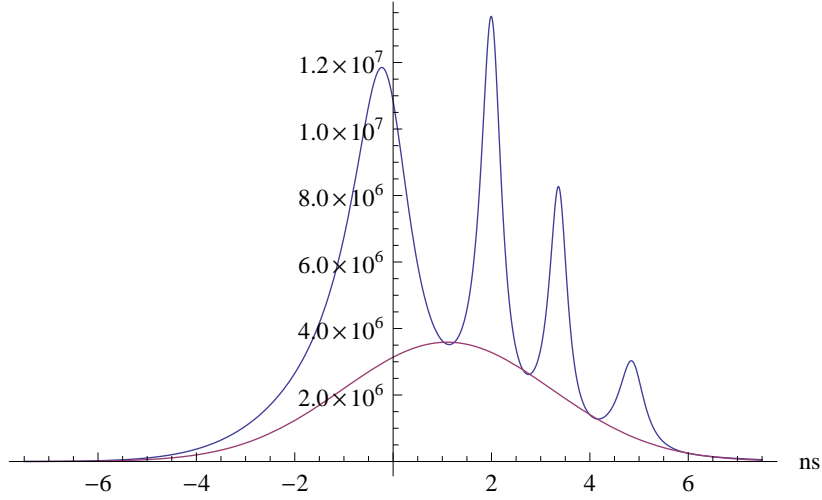}} \hspace*{\fill}
   \caption{Multiply split pulses in medium over the entire region
   of incident pulse.
Number density $10^{17}$cm$^{-3}$, target length = 3 cm, laser
      of power $\sim 4 \times 2\pi$ pulse,
   duration 3 ns assumed. The initial angle $\theta_0=\pi/4$.
}
   \label{Pulse splitting}
 \end{center} 
\end{figure*}

The most general form of solutions is given in Appendix \lromn2,
eq.(\ref{general sol 1}) $\sim$ eq.(\ref{general sol 3}).
They are written in terms of an angle factor $\theta_0$
and the same area function $\tilde{\theta}(t-x)$ as eq.(\ref{area function}).
The angle $\theta_0$ describes the initial state of targets.
The limit $\theta_0 \rightarrow 0$ or $\pi$,
describing the fully excited or the fully ground state, gives the solutions above.
A finite value of $\theta_0$ gives a kind of the dark state
\cite{dark state}, a measure of quantum mixture of the
metastable excited and the ground state.

Two important parameters that describe
solutions of eqs.(\ref{amplifier: ground 0}) $\sim$ (\ref{area function}) are
$\alpha$ and the dimensionless constant $K$ in
the area function $\tilde{\theta}$, and they are given by
\begin{eqnarray}
&&
\alpha = 3\pi\sqrt{\frac{\gamma_{je}\gamma_{jg}}{E_{je}^3 E_{jg}^3}}
\frac{E_c \omega n}{E_c^2 - (\omega - E_{eg}/2)^2}
\,,
\label{alpha def}
\\ &&
K \equiv  \tilde{\mu}\delta \epsilon_0^2 \sim 
\frac{\mu_{gg}}{4}\delta \epsilon_0^2
= \frac{3\pi}{2}\frac{\delta \gamma_{jg} \epsilon_0^2}{E_{jg}^2(E_{jg}^2 - \omega^2)}
\,,
\end{eqnarray}
where $\delta$ is the pulse duration.
For the Ba D-state, they are numerically
\begin{eqnarray}
&&
\alpha \sim 0.052
{\rm cm}^{-1} \frac{n}{10^{16} {\rm cm}^{-3}}
\frac{1.53 \omega}{1.53^2 - (\omega - 0.707)^2}
\,,
\\ &&
K \sim 18 \frac{\delta}{{\rm ns}}\frac{{\rm eV}^2}{2.24^2 - \omega^2} 
\frac{\epsilon_0^2}{10^6 {\rm W mm}^{-2}}
\,,
\end{eqnarray}
with $\omega$ to be given in the eV unit.

For targets of short length, we may globally 
characterize the pulse modification in terms of
the gain and the loss.
Relative to the input pulse power (irradiated at $x=0$)
the transmitted pulse at the end of target (placed at $x=L$)
has the gain or loss factor,
\begin{eqnarray}
&&
\frac{1}{1\pm \alpha L \sin \tilde{\theta} + (\alpha L)^2 \sin^2 \tilde{\theta}/2}
\,.
\end{eqnarray}
We may think of sufficiently short pulse so that
remaining polarization and field inside the target is
effectively described by those values at an infinite time.
The area function $\tilde{\theta}(y)$ is then replaced by its value at
time infinity $\tilde{\theta}(\infty) \equiv \theta_{\infty}$, 
hence the input pulse area
\begin{eqnarray}
&&
\theta_{\infty} = \tilde{\mu} \int_{-\infty}^{\infty}dy \epsilon_0^2(y)
\,,
\end{eqnarray}
may be used.
The gain or the loss factor of the transmitted pulse is thus given by
\begin{eqnarray}
&&
G \sim \frac{1}{1\pm \alpha L \sin \theta_{\infty} 
+ (\alpha L)^2 \sin^2 \theta_{\infty}/2}
\,.
\end{eqnarray}
A positive gain $G>1$ requires, for $0 \leq \theta_{\infty}\leq 2\pi$,
\begin{eqnarray}
&&
\alpha L < \mp 2 \cot \frac{\theta_{\infty}}{2}
\,.
\end{eqnarray}

\begin{figure*}[htbp]
 \begin{center}
 \epsfxsize=0.6\textwidth
 \centerline{\epsfbox{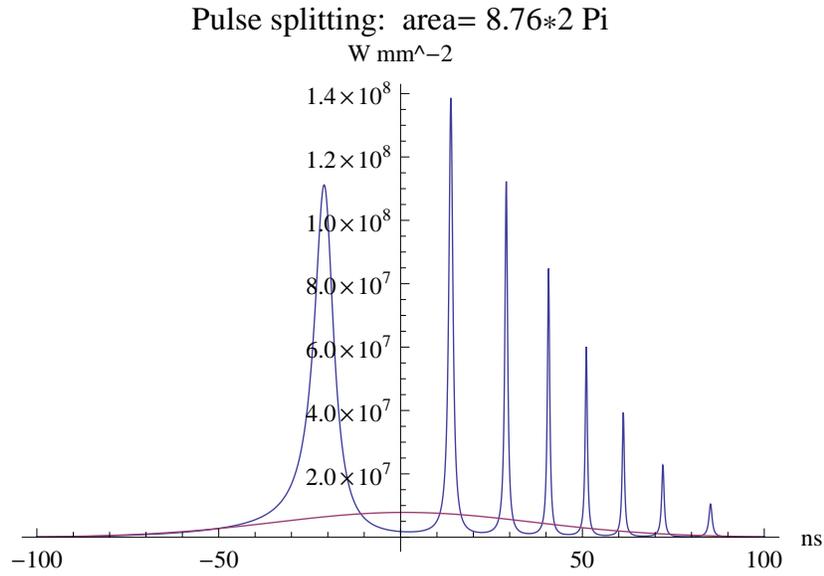}} \hspace*{\fill}
   \caption{Splitting into many pulses in a long target.  
Laser (profile in purple) of intensity corresponding to 8.76 $\times 2 \pi$
area, time duration 50 ns and 5 GHz width is irradiated to a target of
Ba number density $10^{16}$cm$^{-3}$ and length 100 cm. 
}
   \label{many pulse splitting}
 \end{center} 
\end{figure*}

\begin{figure*}[htbp]
 \begin{center}
 \epsfxsize=0.6\textwidth
 \centerline{\epsfbox{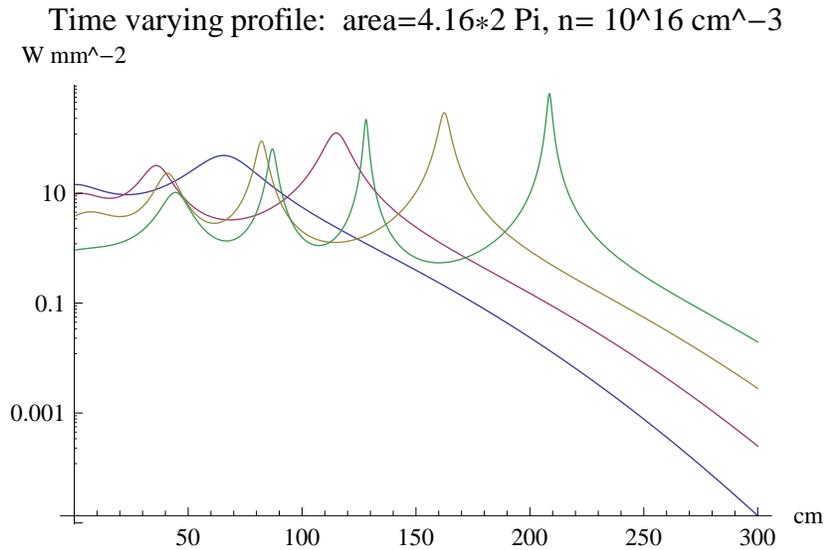}} \hspace*{\fill}
   \caption{Time evolution of field profile at 4 different times
   (blue $\rightarrow$ purple $\rightarrow$ brown $\rightarrow$ green in the time sequence)
   spaced by 0.5 $\times$ 3 ns (pulse duration), $\theta_0=\pi/4$.  Input area close to
$4 \times 2 \pi$ is taken. 
}
   \label{Time evolving field profile}
 \end{center} 
\end{figure*}

\begin{figure*}[htbp]
 \begin{center}
 \epsfxsize=0.6\textwidth
 \centerline{\epsfbox{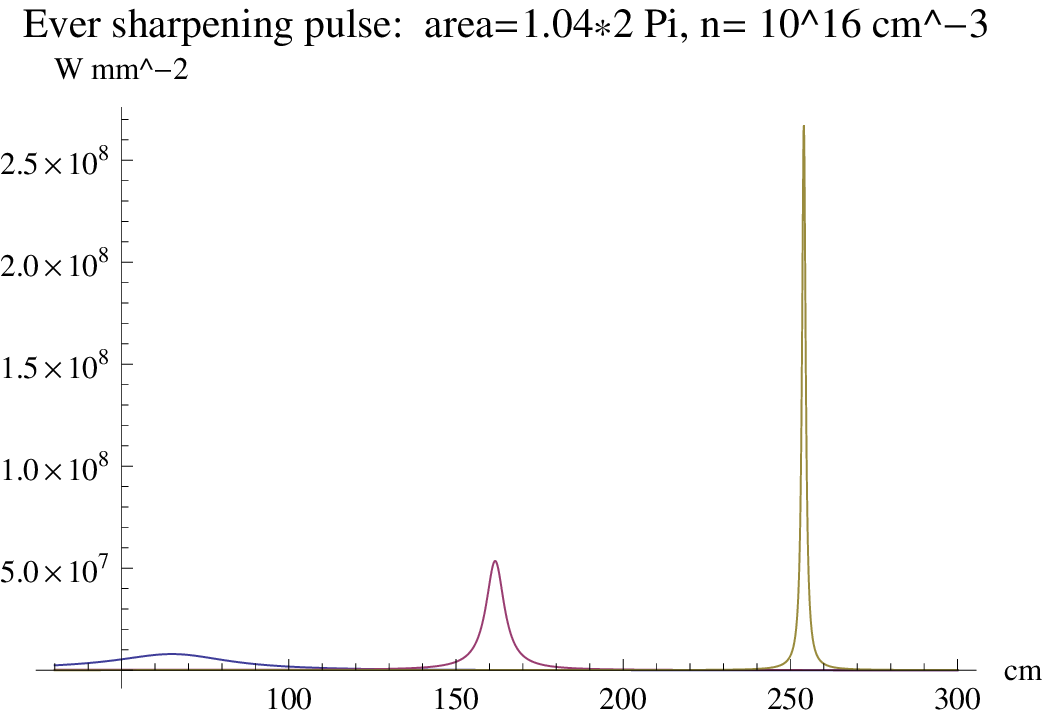}} \hspace*{\fill}
   \caption{Ever sharpening pulse, at 3, 6, 9 ns after the pulse input.
   }
   \label{increasing compression}
 \end{center} 
\end{figure*}

For targets of longer length, the situation is
more complicated:
the pulse splitting occurs along with
compression, as illustrated in Fig(\ref{Pulse compression})$\sim$ 
Fig(\ref{increasing compression}).
Both of these phenomena may be interpreted as signatures
of instability of propagating pulses in medium.
Time profiles of propagating pulses, along with
the incident pulse without medium effects, are shown
in Fig(\ref{Pulse compression}) for the initial area of $\sim 2\pi$
and Fig(\ref{Pulse splitting})
for the area $\sim 8\pi$;
both close to integer multiples of the quantized unit $2\pi$
(the unit becoming important in discussion of solitons below).
One does not observe delayed pulses, which
might have been visible as shifts in two central peaks
of pulses with and without medium.
 This is in contrast to the single photon SR initiated by quantum
fluctuation, where the delay is universally present.
If the target is long enough and the
laser type is close to CW (continuous wave, and not pulse), 
there may be many split pulses
within medium.
In Fig(\ref{many pulse splitting}) we show splitting into 8 
well separated pulses.
The number of separated pulses is $\sim \theta_{\infty}/2\pi$.
Spatial profiles vary with time,
as illustrated in Fig(\ref{Time evolving field profile}).
The ever sharpening pulses are observed, as shown in
Fig(\ref{increasing compression}), but 
in actual situations paired superradiance, not considered here, occurs and
these sharpened pulses are expected to be rounded off by PSR and
become objects close to solitons for which
the pulse shape is unchanged with time.

In summary,
the pulse modification is described by two parameters;
$ \alpha L \,, \theta_{\infty}$.
The larger $\alpha L$ is, the larger modification occurs, while
$\theta_{\infty}$ gives a measure of pulse splitting;
the number of split pulses.
In the Ba example,
$ \alpha L \sim 0.024 n L/10^{16}{\rm cm}^{-2}$ and $\theta_{\infty}
\sim 4 (\delta/{\rm ns}) (\epsilon_0^2/10^6 {\rm W mm}^{-2})$,
both at $\omega = E_{eg}/2$.

We now discuss interaction of two pulses, assuming that both pulses
propagate in the same direction.
The calculation is possible because we solved the one-mode basic equation
in terms of arbitrary initial data which can be
a sum of two pulses of well separated envelopes
at initial times, one of them catching up the other.
The influence of 2nd pulse on 1st pulse
is determined by
using the area sum of two pulses, 
for the case of $\theta_0=0$,
$\tilde{\theta}_1(t-x) + \tilde{\theta}_2(t-x)$,
which replaces $\tilde{\theta}(t-x) $ in the formula,
eq.(\ref{amplifier: ground 0})$\sim$ (\ref{absorber population}).
Suppose that two pulses are well separated by a
distance $X$.
Near the center of 1st pulse, we may take $t-x \sim 0$,
then the effect of 2nd pulse is given by a pulse intensity
modification factor $A$ (to be multiplied by
the energy density $\epsilon_1^2(0)$; the incident pulse
intensity).
This factor may be interpreted 
to define an
effective potential $V$ by using the relation $A = 1-V(X)$.
This is like defining the potential $V$ of 
relativistic particle from the Lagrangian $L$, using
$L = - mc^2 \sqrt{1-\beta^2}- V$. 
For $\theta_0=0$ the potential becomes of the form 
($\mp$ corresponding to amplifier/absorber),
\begin{eqnarray}
&&
V(X) = 1
- \frac{1}{1 \mp \alpha x \sin (\tilde{\theta}_1(0)+\tilde{\theta}_2(X)) 
+ (\alpha x)^2 \sin^2 \left((\tilde{\theta}_1(0)+\tilde{\theta}_2(X))/2\right)}
\,,
\label{two pulse interaction}
\end{eqnarray}
where $x$ is the distance 1st pulse propagated in medium
(2nd pulse is at $x-X$ in medium).
This formula is of generic validity, and we shall apply it
to the problem of two soliton interaction in the next section.

\vspace{1cm}

{\bf \lromn5 \;
Soliton solutions
}

Existence of stable solitons is
anticipated from the topological reason when
the pulse area takes quantized values of an integer times $2\pi$.
This situation is similar, but not identical, to  the self-induced transparancy (SIT)
in the two level system \cite{coherent light propagation in 2 level}.
In our effective $\Lambda-$model
the fictitious pendulum may start from the top $(\theta = 0)$ or the
bottom $(\theta = \pi)$, and come back to the same initial location,
since its motion is limited to a single $\theta-$island in Fig(\ref{theta velocity})
between two end points (where velocity vanishes).
Calculation of the pulse area for this
motion gives the quantized  unit of $2\pi$.

A more physical reason for existence of solitons
is that the propagating pulse in medium
is not stable 
and becomes reshaped via spliting and compression.
The sharpening process however does not last
indefinitely, since the field energy density
increases without bound.
If the coherent region is maintained long enough,
the sharpened pulse eventually emits PSR, the only
possible process in our model, thus
becoming stable against PSR by formation of solitons.
This picture shall be supported in conjunction
with PSR rate computation in the next section.

Solitons in our effective $\Lambda-$model are solutions of
\begin{eqnarray}
&&
\partial_{\eta}\theta(\eta\,, \tau)  = 
\mp \alpha (\cos \theta(\eta\,, \tau) - 1 )
\,, \hspace{0.5cm}
\tau = t - x
\,, \hspace{0.5cm}
\eta = x
\,,
\label{theta eq}
\end{eqnarray}
with the boundary condition suitablle to the
kink solution,
\begin{eqnarray}
&&
\theta(\eta, - \infty) = 0 
\,, \hspace{0.5cm}
\theta(\eta,  \infty) = 2\pi 
\,,
\end{eqnarray}
and a finite energy condition,
\begin{eqnarray}
&&
\int_{-\infty}^{\infty}dt \left(\partial_t \theta (x,t)\right)^2 < \infty
\,.
\end{eqnarray}
-(+) case in eq.(\ref{theta eq}) corresponds to amplifier(absorber).
The anti-kink, or anti-soliton, does not exist,
because the relevant condition,
$\theta(\eta, - \infty) = 2\pi
\,, 
\theta(\eta,  \infty) = 0$
implying $\partial_t \theta < 0$ at some time $t$, is excluded from the 
required positivity of field energy due to
$|E_0^2| \propto \partial_t \theta$.

The explicit form of the tipping angle $\theta$
and the pulse strength $|E^2|$ of soliton solutions is
given in terms of a single function $T(y)$,
\begin{eqnarray}
&&
\theta(x\,, t) = \pm \arccos 
\frac{1 - \alpha^2 (x-T(t-x))^2}{1 + \alpha^2 (x-T(t-x))^2}
\,,
\\ &&
\frac{\mu_{gg}}{4} |E^2(x\,, t)| = 
\pm  \frac{2\alpha \partial_{t} T}{1 + \alpha^2 (x-T(t-x))^2}
\,.
\label{soliton general}
\end{eqnarray}
Finiteness of energy requires a behavior of
the function $T(y)$ towards $\pm$ infinite time,
\begin{eqnarray}
&&
\int^{\infty}dy \frac{|T'(y)|}{T^2(y)} < \infty
\,, \hspace{0.5cm}
\int_{-\infty}dy \frac{|T'(y)|}{T^2(y)} < \infty
\,.
\label{finiteness of soliton energy}
\end{eqnarray}
The function $T(y)$ must be monotonic to give
a positive definite flux.

A simple, linear choice of $T(\tau) \propto \tau$
gives
\begin{eqnarray}
&&
\hspace*{1cm}
|E^2(x, t)| = \frac{2\beta v(1-v)}{(1-v)^2 + \alpha^2 (x-vt)^2}
\,,
\\ &&
\beta = 2\sqrt{\frac{\gamma_{je}}{\gamma_{jg}}}
\frac{E_{jg}^2 (E_{jg}^2-\omega^2)E_c \omega n}
{\sqrt{E_{je}^3 E_{jg}^3}(E_c^2 - (\omega - E_{eg}/2)^2)}
\,,
\end{eqnarray}
where $\alpha$ is defined by (\ref{alpha def}).
The parameter $v$ restricted to $0< v< 1$ here may be regarded as the velocity
of soliton.
In the $v\rightarrow 1$ limit, soliton becomes sharpened without bound;
its width $\propto 1-v$
and its peak value $\propto 1/(1-v)$.
The integrated soliton flux is 
\begin{eqnarray}
&&
\frac{2\pi v\beta}{\alpha} = \frac{4}{3}v \frac{E_{jg}^2 (E_{jg}^2-E_{eg}^2/4)}
{\gamma_{jg}} \times v
\sim 4.75 \times 10^{7} {\rm W mm}^{-2} {\rm cm} \times v
\,,
\end{eqnarray}
(numerical value for Ba),
which is independent of the target number density.

Introduction of higher order powers in $T(\tau)$, like
$\tau^3$, gives asymmetric distortion of pulses.

\begin{figure*}[htbp]
 \begin{center}
 \epsfxsize=0.6\textwidth
 \centerline{\epsfbox{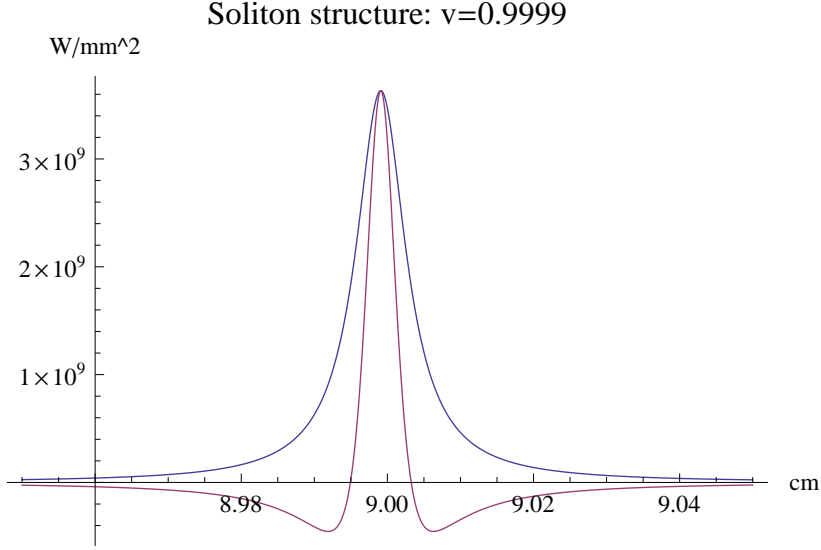}} \hspace*{\fill}
   \caption{Soliton sandwitch structure: field power
and net emission rate, namely, 2 photon (emission - absorption) rate. 
}
   \label{soliton structure}
 \end{center} 
\end{figure*}

\begin{figure*}[htbp]
 \begin{center}
 \epsfxsize=0.6\textwidth
 \centerline{\epsfbox{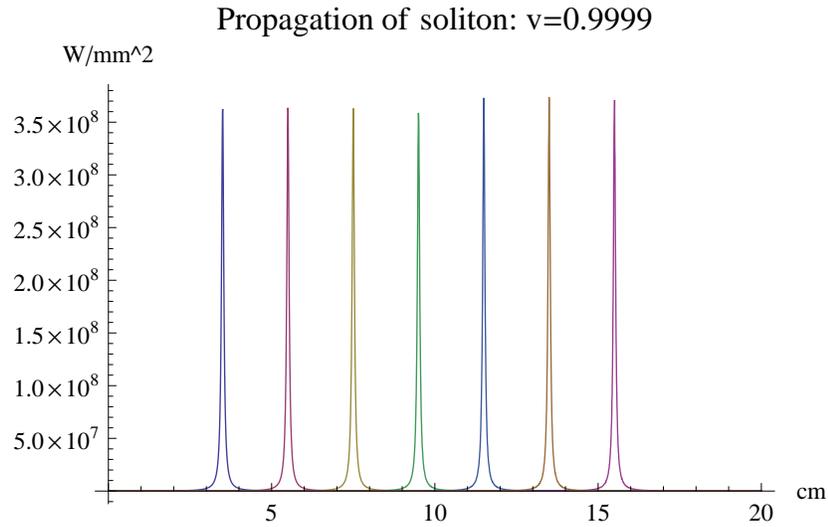}} \hspace*{\fill}
   \caption{Spatial profile invariance of propagating soliton, shown 
at equally spaced different times (the first blue one giving the original pulse).
Ba number density $10^{15}$cm$^{-3}$ and the velocity $0.9999 \times$ 
the light velocity are assumed.
}
   \label{soliton propagation}
 \end{center} 
\end{figure*}

\begin{figure*}[htbp]
 \begin{center}
 \epsfxsize=0.6\textwidth
 \centerline{\epsfbox{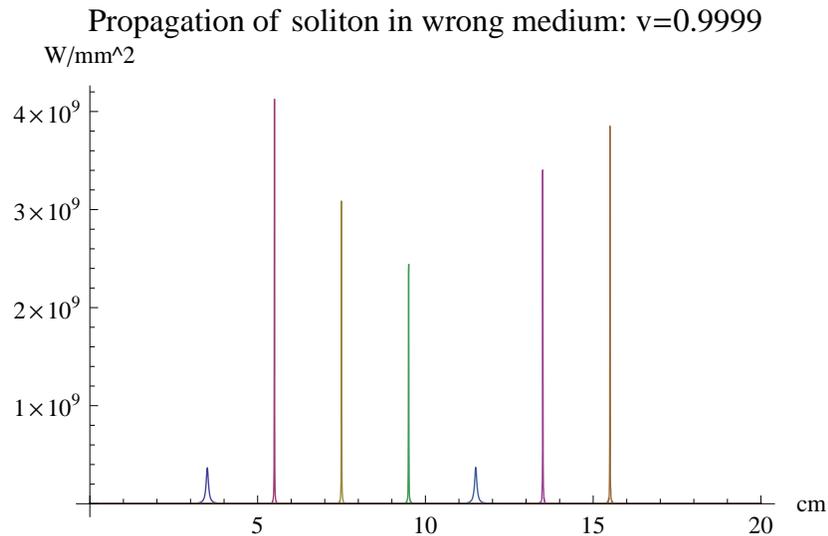}} \hspace*{\fill}
   \caption{Spatial profiles at different times when
the same soliton as in Fig(\ref{soliton propagation}) is put into a medium
   of a mismatched number density 10 times larger than  in Fig(\ref{soliton propagation})
}
   \label{mismatched medium}
 \end{center} 
\end{figure*}

Fig(\ref{soliton structure}) shows
an example of soliton structure within target.
The soliton has a sandwitch structure:
2 photon emission region in the middle is surrounded by
two sides of absorbion region.
This way the net two photon emission seen outside the
target vanishes, as more fully explained in the next section
on PSR rate calculation.
These solitons are stable, keeping their pulse area of $2\pi$
and their shape, as illustrated in Fig(\ref{soliton propagation}).
The soliton size is characterized by $1/\alpha$,
which is $\propto 1/n$ (inversely proportional to the target number density).
When the soliton propagates within medium of a different, hence
mismatched number
density, its shape changes, as illustrated in Fig(\ref{mismatched medium}).

It should be kept in mind that
solitons represent, not only the field but also, the entire 
coherent state of atoms and fields as a whole.
When a soliton exits from a target end
into another region of different environment
(for example, of different target number density or
of different matter including vacuum),
a mismatch of soliton parameters occurs and
the soliton becomes destabilized and necessarily emits PSR.
This gives a simple principle of detecting a
soliton and PSR at the same time.
An obvious obstacle against this
is a fast relaxation process.

\begin{figure*}[htbp]
 \begin{center}
 \epsfxsize=0.6\textwidth
 \centerline{\epsfbox{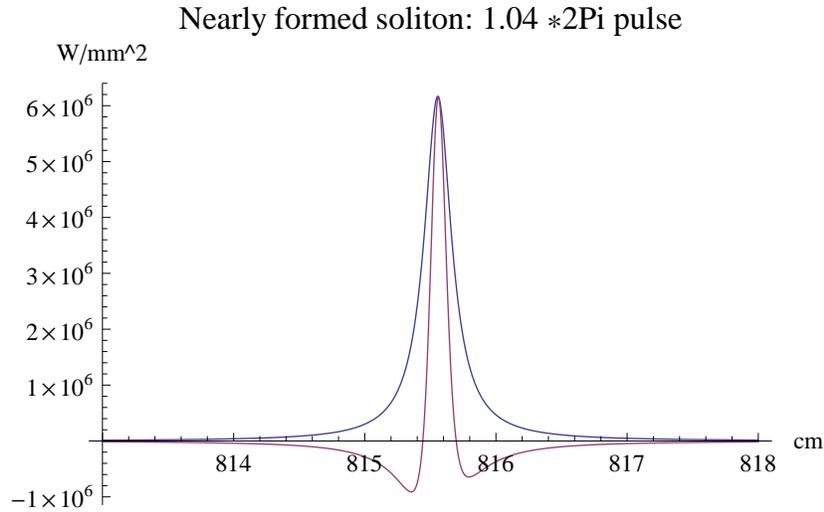}} \hspace*{\fill}
   \caption{Spatial profile of nearly formed soliton.
   The net emission rate and the pulse profile after a long
passage in the ground state amplifier are shown.
$\theta_0 = 0.9999 \times  \pi$, and
Ba number density $10^{16}$cm$^{-3}$ assumed. 
}
   \label{nearly formed soliton}
 \end{center} 
\end{figure*}

We now discuss how solitons might be 
dynamically created in a long target.
The first step towards multiple soliton formation 
is to irradiate a strong pulse of area close to an integer $n \times 2 \pi$
and create $n$ number of well separated pulses.
Each of these pulses are candidates of soliton of area $2 \pi$, but
they must be reshaped.
A method of reshaping would be to recast these pulses into
an amplifier close to the ground state given by the angle $\theta_0 = \pi$.
The reason for this is that at this angle value
the solution is given by eq.(\ref{amplifier: ground 0}), thus
at late times one gets
\begin{eqnarray}
&&
|E_0^2(x,t)| \sim \frac{\epsilon_0^2(t-x)}{1 - \alpha x \tilde{\delta}(t-x) + (\alpha x)^2}
\,,
\end{eqnarray}
where the function $\tilde{\delta}(y)$ is the area - $2\pi$,
and $\tilde{\delta}(y) = \tilde{\theta}(y) - 2\pi \sim
(\partial_y \tilde{\theta})_0 \,y$.
This has the same form as the soliton solution, eq.(\ref{soliton general}),
with 
$T(y) = \tilde{\delta}/(2 \alpha )\sim y (\partial_y \tilde{\theta})_0/(2 \alpha )$,
if $\tilde{\delta}$ is small.
In Fig(\ref{nearly formed soliton}) we illustrate a profile
of nearly formed soliton, constructed by
propagating a nearly $2\pi$ pulse
in a long medium of $\theta_0 \sim \pi$.
The pulse profile constructed this way is similar to, but a little bit 
distorted from,
the soliton profile in Fig(\ref{soliton structure}).
The reshaping becomes perfect when the process of PSR occurs;
the subject of the next section.

\begin{figure*}[htbp]
 \begin{center}
 \epsfxsize=0.6\textwidth
 \centerline{\epsfbox{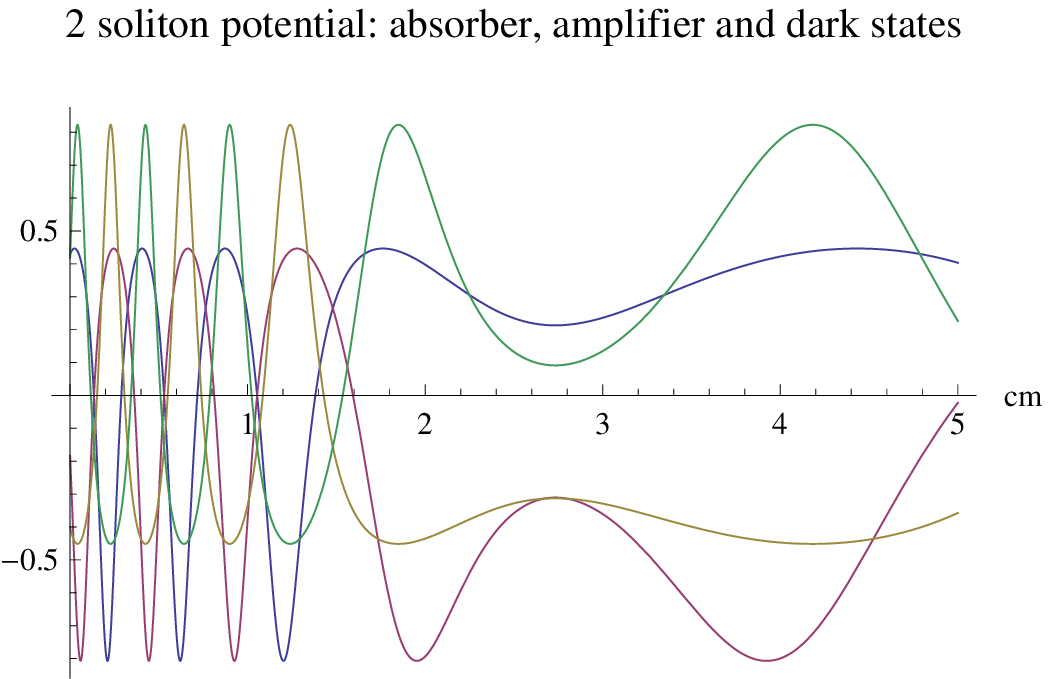}} \hspace*{\fill}
   \caption{Effective potential between two solitons, both located in 
a single medium.
   1st soliton of velocity $0.9 \times$ the
light velocity is at $5$cm away from the target end,
and separated from 2nd of velocity $0.99 \times$ c by
   indicated distance.
Ba number density $5 \times 10^{16}$cm$^{-3}$ is assumed, and
4 different target states, the amplifier  (in blue),
the absorber (purple), and two dark states (in
amplifier and absorber
corresponding to two signs in eqs.(\ref{general sol 1}) and (\ref{general sol 3})) of $\theta_0 = \pi/2$ 
(green and brown) are compared. 
}
   \label{two soliton potential 1}
 \end{center} 
\end{figure*}

Two solitons infinitely separated from each other
in medium
propagate independently, but at finite distances
they start to influence each other.
Two soliton interaction of this kind is similar to
the van der Waals interaction between two neutral atoms
caused by induced dipoles.
Interaction of two solitons may be calculated using
the effective potential given by eq.(\ref{two pulse interaction})
in the preceding section.
Examples of two soliton potential are shown in Fig(\ref{two soliton potential 1})
for various target states.
The conclusion on two soliton interaction
is that its potential has many extremal equilibrium points
for all types of target states. 
As the target number density increases,
the depth of potential (local) minimum becomes larger
and its  locations becomes closer to another soliton.
Implications on this force nature shall be
discussed elsewhere.

\vspace{1cm}

{\bf \lromn6 \;
Theory of PSR and its relation to soliton formation
}

At the outset we would like to point out that
there are two kinds of PSR, the one initiated by a trigger
and another that occurs without (or with a very weak) trigger,
looking like a spontaneous emission at a
superficial level.
These may be regarded as two parts of the coherent two photon
emission under an incident trigger pulse;
the stimulated and the spontaneous parts.
The stimulated part has a rate $\propto N_{\gamma}$
(the number of photons in the incident pulse),
and the spontaneous part $\propto N_{\gamma}^0$.
The rates of these two processes 
differ by a large factor, 
roughly reflecting their available number of photons;
$ O[|E_0(\omega)|^2/\omega]$ (the number within the trigger pulse) 
for the triggered PSR and
$O[\omega^3/(2\pi^2)]$ (the number within a single photon phase space) 
for the trigger-less PSR.
The spontaneous to the stimulated PSR rate
ratio is numerically
\begin{eqnarray}
&&
O[10^{-4}](\frac{\omega}{{\rm eV}})^4 \frac{10^6 {\rm W mm}^{-2}}{|E_0(\omega)|^2}
\,.
\end{eqnarray}

Although its rate is smaller,
there are  merits for PSR without (or with a very weak) trigger;
easiness of two photon simultaneous detection
and the presence of the  PSR time delay,
which can be used as an experimental identification method
for a good PSR event.
It should further be noted that
even a very weak trigger pulse for PSR is very useful
to expedite the target coherence.
An optimal and careful choice of the trigger power is obviously required.

We now discuss  PSR with trigger of a larger rate.
The backward photon emission in
PSR is treated as perturbation to the
 pulse propagation.
The propagation part was analyzed by Maxwell-Bloch equation
and already solved.
We are regarding the process  as a whole single event,
starting from the target triggering,
coherence evolution, until PSR occurs.

Introduce a general perturbation to
the propagation problem, given by a Hamiltonian ${\cal H}_i$
to the effective $\Lambda-$system, which may cause a variety of transitions
depending on the choice of ${\cal H}_i$;
in the case of PSR
${\cal H}_i = \vec{d}\cdot \vec{\epsilon}$ 
where $\vec{\epsilon}$ is the field of a single photon emitted in
the backward direction to the pulse.
For RNPE, ${\cal H}_i \sim G_F \nu_i^{\dagger}\vec{\sigma}\nu_j
\cdot \vec{S}_e$ ($\vec{S}_e$ the electron spin operator),
with $\nu_i$ the neutrino field of a $i$-th mass eigenstate. 

The interaction Hamiltonian gives rise to a
perturbed amplitude of the upper level
in equation for $dc_e/dt$, which contains
the amplitude of upper level $|j \rangle$, 
(see eq.(\ref{upper level amp}) in Appendix \lromn1).
Amplitudes we need to consider in this case
are the unperturbed amplitude $c_j^{(0)}(t)$ and the pertubed amplitude
$\delta c_j(t)$, given respectively by
\begin{eqnarray}
&&
c_j^{(0)}(t) \sim
\frac{d_{jg} }{\omega - E_{jg}}e^{-i(\omega - E_{jg})t}E^*c_g
 - \frac{d_{je} }{\omega + E_{je}}e^{i(\omega + E_{je})t}Ec_e
\,,
\\ &&
\delta c_j(t)
\sim  \frac{d_{jg}}{\omega' - E_{jg}}e^{-i(\omega' - E_{jg})t}e^*c_g 
 - \frac{d_{je} }{\omega' + E_{je}}e^{i(\omega' + E_{je})t}ec_e
\,.
\end{eqnarray}
Pulse field $\times$ perturbed backward field contain spatial
functions, $E \propto e^{-ikx}$ and
$e \propto e^{-ik'x}$,
where $(\omega, k) \,, (\omega', k')$ are frequency and wave number sets of respective waves.

A crucial observation is that at the half  of two level energy difference,
oscillating terms have a common phase,
since
\begin{eqnarray}
&&
-(\omega + E_{je}) = \omega - E_{jg}
\,,
\end{eqnarray}
at $\omega = E_{eg}/2$, which much simplifies the amplitude.

We introduce a new notation for propagating and backward emitted photon
fields by
\begin{eqnarray}
&&
E_0^* e^{i(\omega t - kx)} + E_0 e^{-i(\omega t - kx)}
\,,
\\ &&
e_0^* e^{i(\omega' t - k' x)} + e_0 e^{-i(\omega' t - k' x)}
\,,
\end{eqnarray}
and derive the amplitude of $| j\rangle$ -level,
$c_j^{(0)}(t)  + \delta c_j(t)$, 
by time integration, leading in the Markov approximation to 
\begin{eqnarray}
&&
c_j^{(0)}(t)  = - \frac{e^{iE_c t}}{E_c}
\left( 
e^{i(\omega - \omega_0)t - ikx} E_0^* d_{je} c_e(t) 
+ e^{-i(\omega - \omega_0)t + ikx} E_0 d_{jg} c_g(t) 
\right)
\,,
\\ &&
\delta c_j(t) = - \frac{e^{iE_c t}}{E_c}
\left( 
e^{i(\omega' - \omega_0)t - ik' x} e_0^* d_{je} c_e(t) 
+ e^{-i(\omega' - \omega_0)t + ik' x} e_0 d_{jg} c_g(t) 
\right)
\,,
\end{eqnarray}
where $\omega_0 \approx E_{eg}/2$ is the frequency of triggering pulse.
Neglected terms are rapidly oscillating with $x$ at $k' \sim - k$ 
(giving two counter-propagating waves).

This gives corresponding 
perturbation amplitudes,
\begin{eqnarray}
&&
\delta c_e(t) = i c_g(t)e^{i(k+k')x} \frac{d_{ej}d_{gj}E_0 e_0}{E_c}
\frac{4 \sin (\omega + \omega' - 2\omega_0)t/2}
{\omega + \omega' - 2\omega_0}
\,,
\label{absoption amp}
\\ &&
\delta c_g(t) = i c_e(t)e^{-i(k+k')x} \frac{d_{ej}d_{gj}E_0^* e_0^*}{E_c}
\frac{4 \sin (\omega + \omega' - 2\omega_0)t/2}
{\omega + \omega' - 2\omega_0}
\,.
\label{emission amp}
\end{eqnarray}
One may interpret $e_0$ and $e_0^*$
as annihilation and creation operators of backward emitted photons
according to quantum field theory.

Equation (\ref{emission amp}) describes paired emission of two photons,
counting the stimulated part of amplitude
 ($\propto \sqrt{N_{\gamma}}$) of incident pulse,
and (\ref{absoption amp}) describes an
associated process of paired annihilation of two photons.
The reason why the paired annihilation occurs is that
in medium under propagating field some atoms may be in the
ground state surrounded by ambient two modes of photons.
This amplitude is $\delta c_g$, the perturbed ground state amplitude.

When one computes the probability of PSR,
one first sums amplitudes including $e^{\pm i (k+k')x}$
over all atoms in a coherent medium.
For this purpose we introduce local atomic amplitudes $c_i(x,t)\,, i = e,g$
and divide the entire coherent region $0< x<L$ into cells of size $\Delta x$.
We require $1/k \ll \Delta x \ll L$ for the cell size. 
Under SVEA, namely the
assumption of slow variation of local amplitudes over
the wavelength scale, the amplitude summation  within a cell gives
\begin{eqnarray}
&&
\int_x^{x+\Delta x} dy c_i(y,t) e^{\pm i (k+k')y} 
\approx
c_i(x + \frac{\Delta x}{2}\,, t)e^{\pm i (k+k')\Delta x/2}
\frac{2\sin(k+k')\Delta x/2}{k+k'}
\,.
\end{eqnarray}
The last factor, when squared, gives 
the Dirac delta function in the form,
$\Delta x 2\pi \delta (k+k')$, in the large
$\Delta x $ limit (of $\Delta x \gg$ the wavelength).
The factor $\Delta x $ in front is cancelled by the amplitude squared 
$|e_0|^2$ of a single backward photon.

The factor $2\pi \delta (k+k')$ implies the momentum conservation
working,
independently of the actual finite, but a large  $\Delta x$.
Our basic ansatz in the spirit of SVEA is
to take for $\Delta x$ its largest possible value,
the target size $L$.
Furthermore,
we replace for  ease of numerical computations the fast oscillating function
by a more smoothly varying, yet globally correct, function;
\begin{eqnarray}
&&
\left(\frac{\sin (k+k')L/2}{k+k'} \right)^2
\rightarrow
\frac{1}{(k+k')^2 + 4/(L^2)}
\,.
\end{eqnarray}
We may, somewhat arbitrarily but realistically,
change in rate estimates the factor $L$ here by its fraction, which
gives larger rates than given here.
We may thus regard our procedure of the
replacement $\Delta x \rightarrow L$ as an underestimate of 
more realistic rates.
How much rates are underestimated actually, however,
must be verified by a more laborious numerical
simulation,
assuming discretized sites of atoms.

Rate of the net emission $\propto R_3$ has two types of contributions;
emission for $R_3 > 0$ from atoms populated  more in the excited state, 
and absorption  for $R_3 < 0$ from atoms populated more in the ground state.
The absorption that occurs within target medium
cannot be experimentally measured, and one
measures the positive emission rate at both target ends.
(If the effective rate becomes negative,
no emission is measured since it means an inward emission into
the target inside.)
The spacetime dependence of net rates is thus given by
\begin{eqnarray}
&&
|E_0 \,e_0|^2 \,n R_3(x,t)I(x,t) 
(4\frac{\sin (\omega + \omega' - 2\omega_0)t/2}{\omega + \omega' - 2\omega_0})^2 
(\frac{d_{ej}d_{gj}}{E_c})^2
\frac{4}{(k+k')^2 + 4/(L^2)} 
\,,
\end{eqnarray}
where the pulse related factor $I(x,t)$ is to be given below in eq.(\ref{pulse related factor}).
We may define the net spectral rate (net probability per unit time
which can also become negative) as
\begin{eqnarray}
&&
|E_0^2| 4\pi V \omega n R_3(x,t)I(x,t) (\frac{d_{ej}d_{gj}}{E_c})^2 \times
4 \left( \frac{\sin (\omega - \omega')L/2}{\omega - \omega'}\right)^2
\,,
\end{eqnarray}
by the short time average using the well known formula,
\begin{eqnarray}
&&
\lim_{t \rightarrow \infty}\frac{4}{t}\left(
\frac{ \sin (\,(\omega + \omega' - 2\omega_0)t/2\,)
}{\omega + \omega' - 2\omega_0}\right)^2
= 2\pi \delta(\omega + \omega' - 2\omega_0) 
\,.
\end{eqnarray}
The large time limit is valid since oscillation in time
is very fast compared to polarization development
under all practical situations we consider.

Convolution with the frequency distribution of the input pulse is now necessary.
We assume a Gaussian power spectrum of the form,
\begin{eqnarray}
&&
F(\omega\,; \omega_0, \delta)
= \frac{\epsilon_0^2(\omega_0)}{\sqrt{\pi}\delta}
\exp[- \frac{(\omega - \omega_0)^2}{\delta^2}]
\,,
\end{eqnarray}
with the frequency width 
$\delta (= O(50 \sim 1)$GHz for commercially available laser).
The wave number integration, along with the convolution, gives
\begin{eqnarray}
&&
\int_{0}^{\infty}d\omega
\int_{-\infty}^{\infty}dk' 
\frac{F(\omega\,; \omega_0, \delta)}{(\omega +k')^2 + 4/L^2}
\delta(\omega + |k'| - E_{eg})
\sim
\frac{\pi L}{2}F(E_{eg}- \omega'\,; \omega_0, \delta)
\,,
\end{eqnarray}
leading to the differential spectrum 
$d^2\Gamma /d\omega' dt$ (having the dimension of rate\,, 1/time) of the backward photon of
energy $\omega'$;
\begin{eqnarray}
&&
\hspace*{-1cm}
8\pi^2 n^2 V(3\pi)^2 
\frac{\gamma_{je}\gamma_{jg}}{E_c^2 E_{je}^3 E_{jg}^3}
\frac{E_{eg} \pi }{4}
\frac{E_{eg} - \omega'}{(2\omega' - E_{eg})^2 + 4/L^2}
\frac{\epsilon_0^2(\omega_0)\exp[- \frac{(\omega' - E_{eg} + \omega_0)^2}{\delta^2}]}{\sqrt{\pi}\delta}
\frac{\exp[-\frac{(t-x)^2}{\Delta^2}]}{\sqrt{\pi}\Delta}
r_3( x,t\,;\omega') I( x,t\,;\omega')
\,, 
\nonumber \\ &&
\\ && \hspace*{-2cm}
r_3( x,t\,;\omega') = \pm
\frac{1}{\sqrt{1+ \gamma^2}}
\frac{\cosh (\alpha x\sin \theta_0 )\cos \theta_0
(1 - \cos \theta_0 \cos \tilde{\theta})
\pm \sinh (\alpha x \sin \theta_0 ) \cos \theta_0\sin \theta_0\sin \tilde{\theta}
-  (\cos \theta_0 - \cos \tilde{\theta})}
{\cosh (\alpha x \sin \theta_0 ) (1- \cos \theta_0 \cos \tilde{\theta}) 
\pm \sinh (\alpha \sin \theta_0 x) \sin \theta_0\sin \tilde{\theta} - 
\cos \theta_0 (\cos \theta_0 - \cos \tilde{\theta})}
\,,
\nonumber \\ &&
\\ &&
\hspace*{-1cm}
I( x,t\,;\omega') =
\frac{\sin^2 \theta_0}
{\cosh (\alpha x\sin \theta_0 ) (1- \cos \theta_0 \cos \tilde{\theta}) 
\pm \sinh (\alpha x\sin \theta_0 ) \sin \theta_0\sin \tilde{\theta} - 
\cos \theta_0 (\cos \theta_0 - \cos \tilde{\theta})}
\,. 
\label{pulse related factor}
\end{eqnarray}
For the input pulse we assumed a  time structure of the
Gaussian form, characterized by a width $\Delta$ and centered at $t=x$.
Two cases of $\mp$ correspond to the state of targets;
absorber (amplifier).

Note that we used neither the conventional Gaussian width given by the variance,
nor the half width at half its maximum.
Our width is $\sqrt{2} \,\times $ the Gaussian and
$1/\sqrt{\ln 2} \,\times $ the half width.

For the Ba D-state, the basic unit of differential spectrum
for the target number density $n= 10^{16}$cm$^{-3}$ and 
the pulse power $\epsilon_0^2 = 10^6$W mm$^{-2}$ is
(under the $\hbar = 1\,, c=1$ unit)
\begin{eqnarray}
&&
8\pi^2 (3\pi)^2 (10^{16}{\rm cm}^{-3})^2 {\rm cm}^2\,
{\rm GHz\, MHz}\, 10^6{\rm W mm}^{-2} = 5.7 \times 10^{19}{\rm Hz \, eV} 
\,,
\end{eqnarray}
to be multiplied by
\begin{eqnarray}
&&
(\frac{n}{10^{16}{\rm cm}^{-3}})^2\frac{V}{{\rm cm}^3}
\frac{E_{eg} - \omega'}{(2\omega' - E_{eg})^2 + 4/L^2}
\frac{\epsilon_0^2}{10^{6}{\rm W mm}^{-2}}
\frac{\exp[- \frac{(\omega' - E_{eg} + \omega_0)^2}{\delta^2}]
\,d\omega'}{\sqrt{\pi}\delta}
\nonumber \\ &&
\hspace*{1cm}
\times
r_3(\omega'\,, x,t) I(\omega'\,, x,t)\frac{\exp[- \frac{(t-x)^2}{\Delta^2}]\, dt}{\sqrt{\pi}\Delta}  \,,
\end{eqnarray}
where all energies should be given in the eV unit.
Two cases of
+(-) correspond to the absorber (amplifier).

The time dependence of rate
$d^2\Gamma/d\omega' dt$ is due to the rapidly varying pulse shape.
A more practical measure of event rate is
time integrated event number per a shot of pulse;
\begin{eqnarray}
&&
\int_{-\infty}^{\infty}dt \,\frac{d^2\Gamma(\omega'\,;x, t)}{d\omega' dt}
\,.
\end{eqnarray}
From the measurement point of view, only two locations 
are of practical importance; $x = 0$ the target left end
for measurement of the backward emitted photon of PSR, and
$x=L$ the right end for measurement of the forward emitted photon.

The frequency spectrum of backward photon is sharply peaked at the
middle point of level spacing, $\omega' = E_{eg}/2$,
if the trigger frequency is tuned to this value, $\omega_0 = E_{eg}/2$.
A quantity of practical importance  for the overall rate is
the frequency integrated (time dependent) rate given by
\begin{eqnarray}
&&
\int_{0}^{\infty}d\omega' \,\frac{d^2\Gamma(\omega'\,;x, t)}{d\omega' dt}
=
\int_{0}^{\infty}d\omega' 
\,\frac{\exp[- \frac{(\omega' - E_{eg} + \omega_0)^2}{\delta^2}]
\,d\omega'}{\sqrt{\pi}\delta}
\frac{E_{eg} - \omega'}{(2\omega' - E_{eg})^2 + 4/L^2}\,F(\omega'\,; x, t)
\,,
\end{eqnarray}
where $F(\omega'\,; x, t)$ is slowly varying in frequency $\omega'$.
We may approximate this frequency integral by taking correct
behavior in the two limiting regions, $\delta L \gg 1$ and $\delta L \ll 1$, and
smoothly interpolating in the interval. At the tuned point $\omega_0 = E_{eg}/2$, one
may adopt the following approximation,
\begin{eqnarray}
&&
\int_{0}^{\infty}d\omega' 
\,\frac{\exp[- \frac{(\omega' - E_{eg}/2 )^2}{\delta^2}]}{\sqrt{\pi}\delta}
\frac{E_{eg} - \omega'}{(2\omega' - E_{eg})^2 + 4/L^2}\,F(\omega'\,; x, t)
\approx
\frac{\frac{E_{eg}}{2}F(\frac{E_{eg}}{2}\,; x, t) 
\sqrt{\pi}L^2}{\sqrt{\pi} + 2 L\delta}
\,.
\end{eqnarray}
We checked that this approximation is valid to an accuracy of $\sim $ 15\% level.

Assuming that the pulse shape variation is slow,
one may introduce a slowly varying, time dependent rate by
taking the short time-average and frequency-integration, to
obtain a total net  event number (emission - absorption
event) per a shot of pulse
as
\begin{eqnarray}
&&
{\cal N}(x,t) =
3.2 \times 10^{24} (\frac{n}{10^{16}{\rm cm}^{-3}})^2\frac{\epsilon_0^2}{10^6 {\rm W mm}^{-2}}
\frac{V}{{\rm cm}^3}\,r_3(\frac{E_{eg}}{2}, x, t) I(\frac{E_{eg}}{2}, x, t)
\frac{\sqrt{\pi}L/{\rm cm}}{\sqrt{\pi} + 2 L\delta}
\,.
\end{eqnarray}
The last factor for the pulse duration of $\delta = 45$GHz and
$L = 3$ cm is $\sim 0.5$.

\begin{figure*}[htbp]
 \begin{center}
 \epsfxsize=0.6\textwidth
 \centerline{\epsfbox{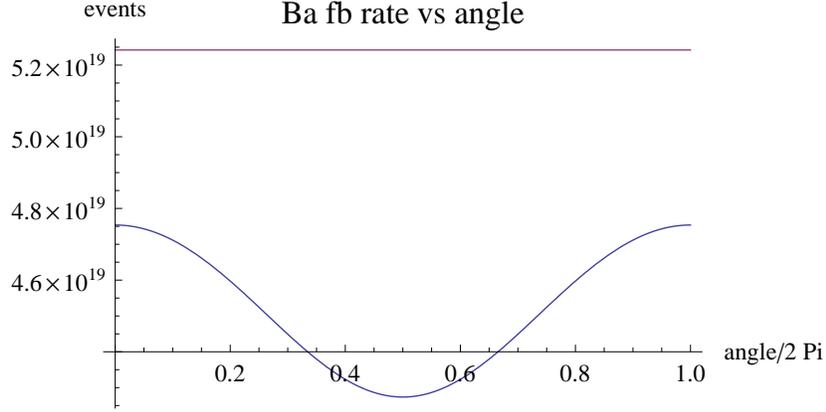}} \hspace*{\fill}
   \caption{ Angle ($\theta_0$) dependence of forward (blue) and backward (purple) rates 
measured at the passage time of the target end (taken 3cm).
Power $10^6$W mm$^{-2}$, number density $10^{16}$cm$^{-3}$, laser
   duration 3 ns assumed.   
}
   \label{angle dependence}
 \end{center} 
\end{figure*}

\begin{figure*}[htbp]
 \begin{center}
 \epsfxsize=0.6\textwidth
 \centerline{\epsfbox{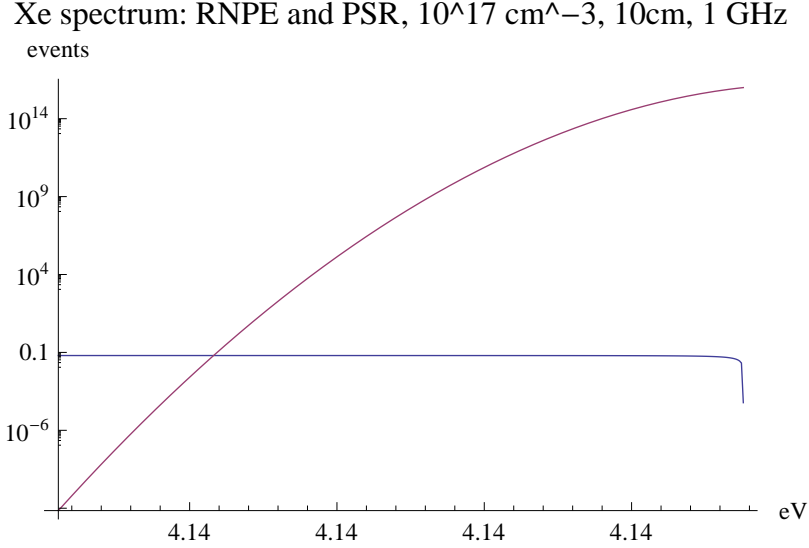}} \hspace*{\fill}
   \caption{Rate of mass 1 meV pair emission (in blue)
and the background PSR rate (in purple) very near the neutrino mass threshold. 
Excited 
Xe number density $10^{17}$cm$^{-3}$ of target length 10 cm
and laser intensity
$10^6$W mm$^{-2}$ of frequency width 1 GHz of time duration 30 ns are assumed.
When the blue curve is above the purple curve, S/N $> 1$.
At the farthest left is the threshold of the pair mass, (8 + 1) meV,
and at the farthest right is that of the pair mass, (1 + 1) meV,
the entire abscissa distance in
this figure being separated by a very small 
energy interval $\sim 4.6 \mu$eV or $\sim $7.1 GHz,
hence their precise values are not shown.
}
   \label{rnpe rate vs n-mass}
 \end{center} 
\end{figure*}

The more effective and useful
event rate (number of events per unit time) in actual experiments is
given by this event number per a shot
divided by the repetition cycle time $\tau_r$
of excitation and trigger.
For instance, if the pulse repetition cycle $\tau_r$ is 1 msec,
a practical experimental rate is ${\cal N}/\tau_r = 10^3 {\cal N} $ Hz.
We caution that
all event numbers shown in Fig(\ref{angle dependence}) $\sim$ 
Fig(\ref{rnpe rate vs n-mass}) 
are events per a shot of pulse, and one has to
multiply $1/\tau_r$ (depending on experimental setup) for
effective and more realistic rates per unit time.

We now exhibit several figures to illustrate physics of analytic results
given here.
These PSR rates are computed,  using the rate formula and analytic
solutions of the pulse propagation for $r_3\,, I_1$.
The first figure Fig(\ref{angle dependence}) shows the initial angle
$\theta_0$ dependence of the backward rate (measured at $x=0$) and 
the forward rate (at $x= L$).
Except at soliton formation discussed later, the forward-backward asymmetry in rate
is not large (typically $\leq 15 \%$).

\begin{figure*}[htbp]
 \begin{center}
 \epsfxsize=0.6\textwidth
 \centerline{\epsfbox{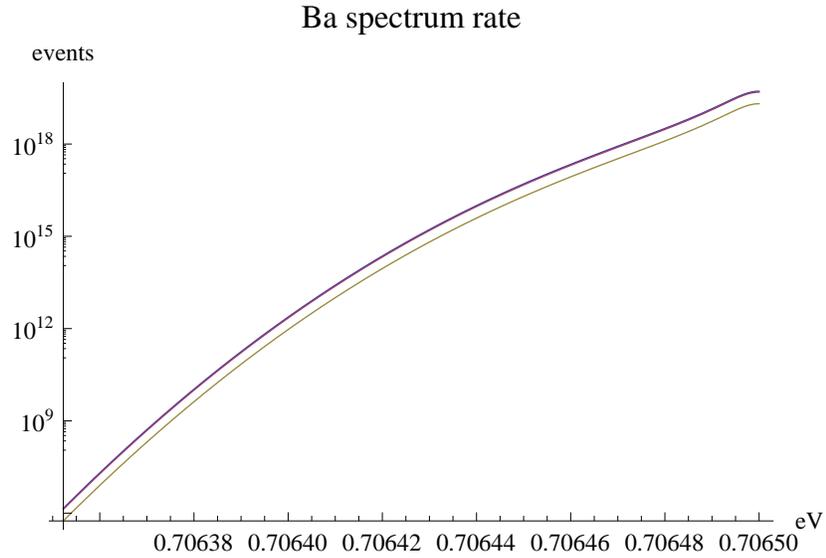}} \hspace*{\fill}
   \caption{Spectrum rate at 3 different times, (0.1, 1, 5) $\times$
 the passing time of the target end.
  Power $10^6$W mm$^{-2}$, number density $10^{16}$cm$^{-3}$, target length = 3 cm, 
laser duration 3 ns, $\theta_0=\pi/4$ assumed.
Rates monotonically decrease in this range of time sequence.
}
   \label{Backward photon energy spectrum for Ba}
 \end{center} 
\end{figure*}

An example of
Ba backward photon spectrum is shown 
in Fig(\ref{Backward photon energy spectrum for Ba}).
In this computation we took the Gaussian frequency distribution
of input laser, its width given by 45 GHz.
Both spectral shape and rate are
indistinguishable up to the passage time
of pulse at the target end, but
the rate rapidly decreases much beyond the passage time.
It thus becomes important to devise a fast
recycling scheme for excitation and trigger of the target,
in order not to wait for a null result.

\begin{figure*}[htbp]
 \begin{center}
 \epsfxsize=0.6\textwidth
 \centerline{\epsfbox{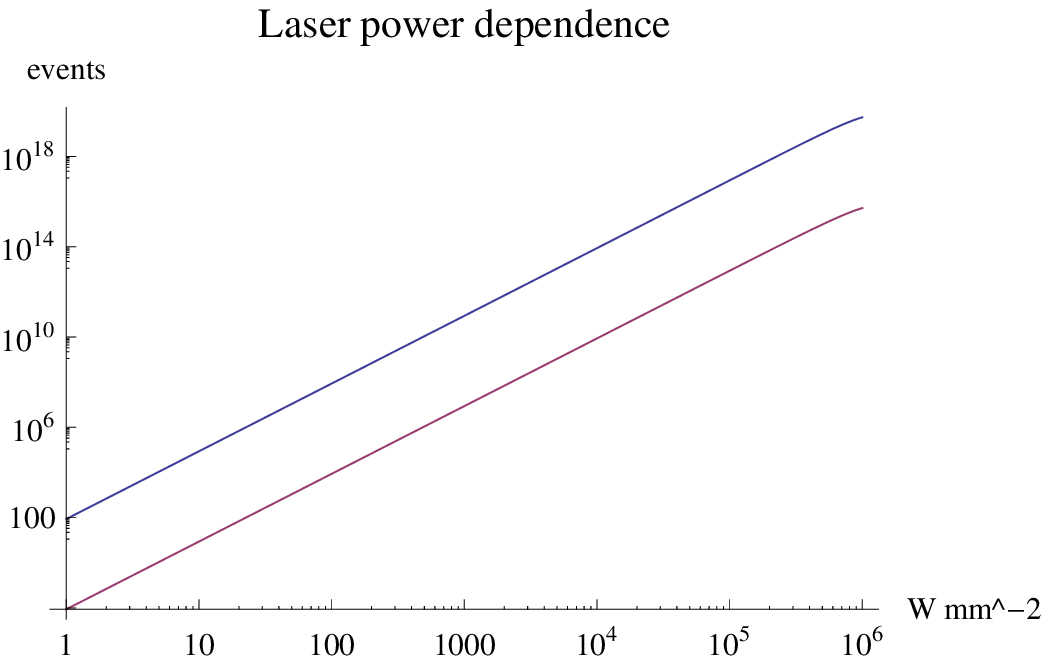}} \hspace*{\fill}
   \caption{Rate at the arrival time of the target
end as a function of laser power.
   Target number denisty $10^{16}$ (in blue), $ 10^{14}$cm$^{-3}$ (in purple), target length = 3 cm, laser
   duration 3 ns of 45 GHz, and $\theta_0=\pi/4$ are assumed.
}
   \label{Effective rate vs power}
 \end{center} 
\end{figure*}

\begin{figure*}[htbp]
 \begin{center}
 \epsfxsize=0.6\textwidth
 \centerline{\epsfbox{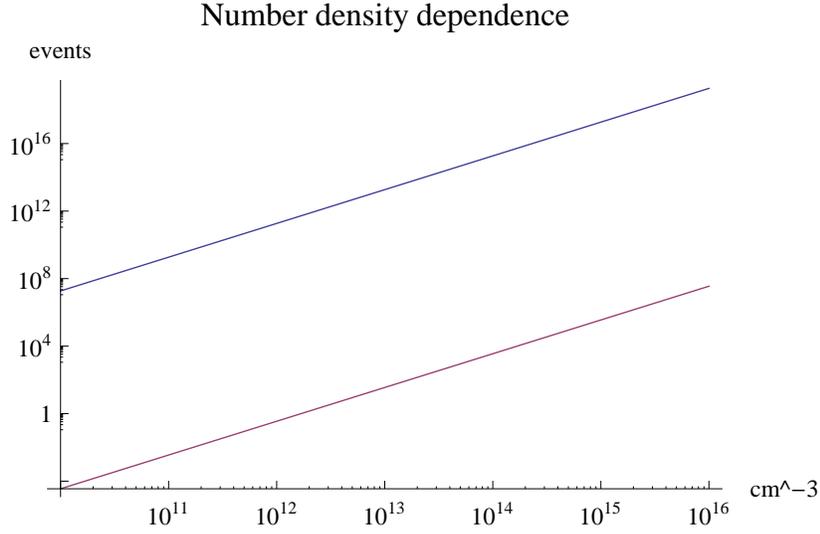}} \hspace*{\fill}
   \caption{Rate at the arrival time of the target
end as a function of Ba number density.
   Two power values $10^6$ (in blue), $ 10^2$W mm$^{-2}$ (in purple), target length = 3 cm, laser
   duration 3 ns, $\theta_0=\pi/4$ assumed.
}
   \label{Effective rate vs number density}
 \end{center} 
\end{figure*}

\begin{figure*}[htbp]
 \begin{center}
 \epsfxsize=0.6\textwidth
 \centerline{\epsfbox{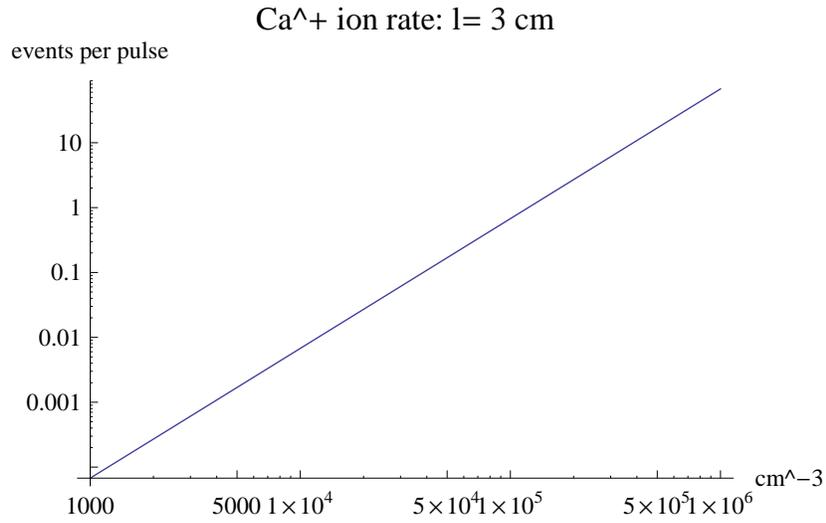}} \hspace*{\fill}
   \caption{Rate for Ca$^+$ ion vs number density.
   Power $10^6$W mm$^{-2}$, target length = 3 cm, laser
   duration 30 ns of 1 GHz width, $\theta_0=\pi/4$ assumed.
}
   \label{ca-ion rate vs number density}
 \end{center} 
\end{figure*}

The laser power dependence of rate is shown in 
Fig(\ref{Effective rate vs power}).
Fig(\ref{Effective rate vs number density})
shows dependence on the target number density dependence,
indicating that PSR may be detectable for Ba number densities
as low as $10^4$cm$^{-3}$ or even less, by taking into account
a fast repetition cycle time of, for instance $O[\mu {\rm  sec}]$
($O[10^6]$ times the event number of 
Fig(\ref{Effective rate vs number density}) for the rate per second),
whose precise value is determined by experimental conditions and not by a
theoretical calculation.
Dependence on the number density is roughly $\propto n^2$.
This may open another interesting possibility of detecting PSR in
alkhali earth ions such as Ca$^+$, Sr$^+$ and Ba$^+$, which
have the $\Lambda-$type level structure.
The actual PSR rates, calculated by using experimentally
known level spacings and decay rates for these ions, are somewhat smaller,
as illustrated in Fig(\ref{ca-ion rate vs number density}) for  Ca$^+$ ion.
Maintaining coherence in trapped ions might however be
easier than in the gaseous phase.
(Even production of the crystalized ion has been achieved.)
A fast repetition cycle of excitation and trigger laser
of order 1 msec might be quite sufficient to obtain 
detectable rates under good envirornment of the ion trap.

One might have a suspicion that
PSR is not detectable, because
a single photon SR rate is always larger than the two photon PSR rate
due to a larger spontaneous emission rate, when both
considered as elementary processes.
This is not always true, because what should be compared in
triggered experiments is the triggered PSR time vs
the delay time of trigger-less SR, and
these two times scale with the target number $N$ as
$\propto N^{-2}$ vs $\propto N^{-1}$.
With a sufficiently large $N$,
the triggered PSR occurs before the delayed SR occurs.

Incidentally, (J=0 $\rightarrow$ J=0 ) PSR transitions, certainly present for
alkhali earth atoms (Sr, Ca etc.) and Yb, are very interesting from the point of
realizing excellent quantum entanglement,
because two emitted PSR photons are well entangled in
their angular momenta, back to back emitted  directions,
and identical photon energy, all to good accuracy.

\begin{figure*}[htbp]
 \begin{center}
 \epsfxsize=0.6\textwidth
 \centerline{\epsfbox{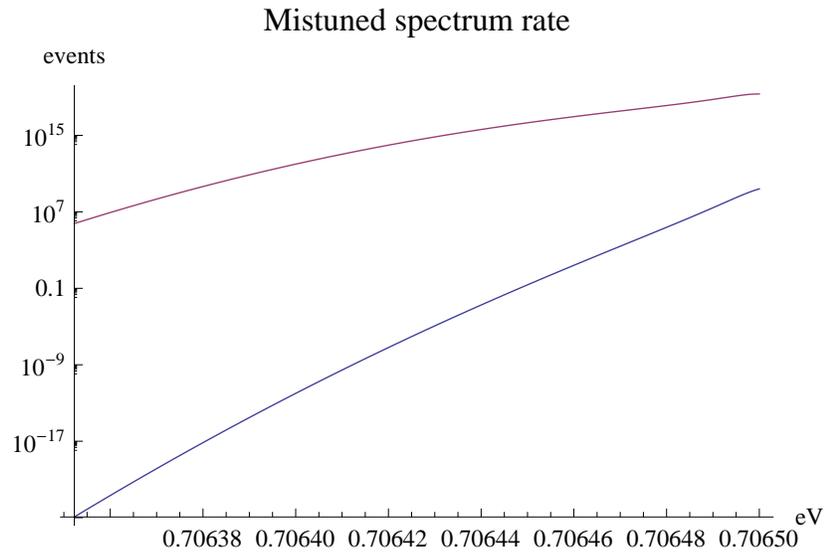}} \hspace*{\fill}
   \caption{  Photon sepctrum in blue when the input laser
   frequency is mistuned by $10^{-4}$ away from the middle energy of $E_{eg}/2$.
   For comparison the original spectrum at the tuned frequency is also shown in purple.
}
   \label{Photon spectrum at mistuned trigger}
 \end{center} 
\end{figure*}

Fig(\ref{Photon spectrum at mistuned trigger}) shows a spectrum rate
at a mistuned frequency of the input laser.
Note that for an infinitely long medium
the momentum and the energy conservation forces PSR
spectrum to have a $\delta-$function like peak at the half energy.
In practice, the target has a finite length
and violation of the momentum conservation
leads to a small tail away from the peak location
of $\omega=E_{eg}/2$.
The amount of suppressed tail contribution
depends much on the frequency distribution
of the tail part of irradiated laser.
A large suppression factor seen here
is due to 
the Gaussian frequency distribution, in this
case of width 45 GHz.
This large suppression 
is encouraging from the point of enhancing the signal to the background ratio
of RNPE/PSR.
We shall have much more to say on this in the last section
when we discuss prospects for RNPE.

We now demonstrate that solitons are stable against
two photon emission.
As discussed above, two photon emission associated with
pulse propagation accompany simultaneously two photon absorption,
since it may be induced by surrounding field.
Thus, the net emission rate is in proportion to $R_3 I$,
the product of population difference and the power of
propagating pulse.
For soliton solutions the
time integrated rate at the target end is
\begin{eqnarray}
&&
\int_{-\infty}^{\infty}dt\, 
\frac{2\alpha T'(t-L) \left(1 - \alpha^2 (L - T(t-L)\,)^2\right)}
{\left(1 + \alpha^2 (L - T(t-L)\,)^2 \right)^2}
\nonumber
\\ &&
= 2 \int_{-\infty}^{\infty}dx\,\frac{1-x^2}{(1+x^2)^2} = 0
\,.
\end{eqnarray}

From Fig(\ref{soliton structure}) one sees that the emission region in the central part
of the target is sandwitched by two absorption regions,
which gives a balanced vanishing net rate.
This result holds at any target point.
In more general target states excluding solitons, the integral of this product 
$\propto \cos \theta \partial_t \theta$
gives a difference,
$\sin \theta (t= \infty) - \sin \theta (t= -\infty)$,
which is non-vanishing.

An ideal method of observing PSR might be
creation of many solitons at the first stage,
and their artificial  destruction by controlled means
for detection of PSR photons at the second stage.

\begin{figure*}[htbp]
 \begin{center}
 \epsfxsize=0.6\textwidth
 \centerline{\epsfbox{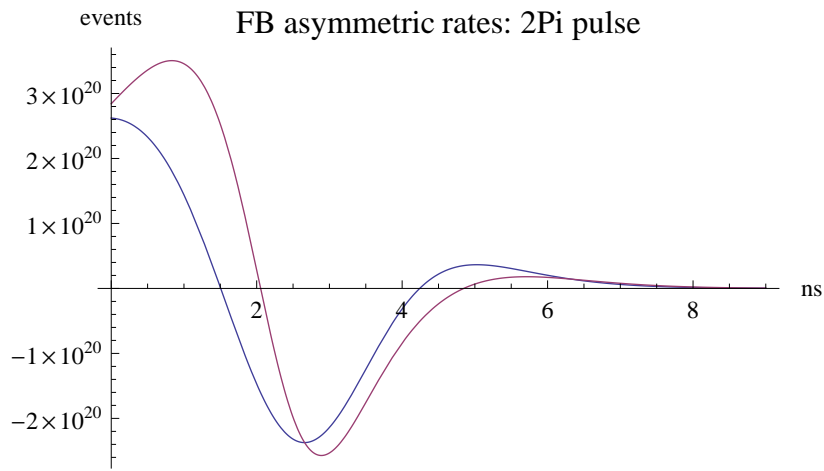}} \hspace*{\fill}
   \caption{Forward-backward asymmetric PSR rates.
   The backward rate in blue is the measured rate at $x=0$, and the forward rate 
in purple at $x = L$
   (the target length 10 cm taken here).
   The rate here is (emission - absorption), hence the negative value region
   gives zero measurement.
$\theta_0 = 0.9999 \times  \pi$, and
Ba number density $10^{16}$cm$^{-3}$ are assumed. 
}
   \label{fb asymmetry}
 \end{center} 
\end{figure*}

The forward-backward asymmetry is expected to be large, both 
immediately prior to soliton formation and immediately after
their destruction.
We plot in Fig(\ref{fb asymmetry}) time evolution of the asymmetric rates
for $\sim 2\pi$ pulse.
This asymmetry may be used to detect PSR itself under
large symmetric backgrounds.

\vspace{1cm}

\vspace{1cm}

%%%%%%%%%%%%%%%%%%%%%%%%%%%%%%%%%%%%%%%%%%%%%%%%%%%

{\bf \lromn7 \; Outlook for radiative neutrino pair emission (RNPE)}

We shall briefly sketch prospects towards  
our goal of the precision neutrino mass spectroscopy, 
by providing rate, spectrum and S/N.
Details of rate computations, including effects of all six
thresholds, mixing angles, Majorana vs Dirac distinction \cite{my-06},
 and Majorana CP phases \cite{pv} 
shall be presented in a separate publication.

Atoms ideal for PSR may not be appropriate
for RNPE detection, due to two reasons:
(1) the RNPE process $|e\rangle \rightarrow |g \rangle
+ \gamma + \nu_i \nu_j \,, i,j = 1,2,3,$
requires a large spin flip amplitude \cite{my-06} 
(similar to M1 transition for
the electronic part of transition matrix element) 
between the intermediate
state and either of the initial or the ground states,
while a large PSR rate may require larger
E1 transition,
(2) PSR photons might become a serious background against RNPE,
which means that smaller PSR rates are better for
RNPE detection.
To the best of our knowledge,
 Xe atom in solid form is among the best
candidate atoms for RNPE.
The candidate metastable state is the first
excited state, 
$ 5p^5(^2P_{3/2})6s ^2[3/2]_2$,
a $J=2$  pair state of electron and hole \cite{pv}.
Other candidates might be metastable states of
spin configuration (of two electrons)
different from the ground state,
commonly seen in diatomic molecules such as $O_2$,
an object worthy of serious consideration.

\begin{figure*}[htbp]
 \begin{center}
 \epsfxsize=0.8\textwidth
 \centerline{\epsfbox{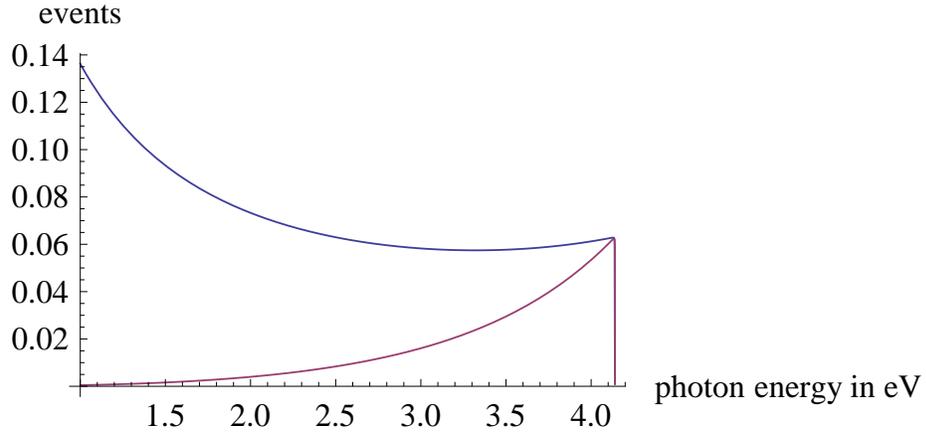}} \hspace*{\fill}
   \caption{RNPE spectrum rate (event number per pulse) starting from the
   neutrino threshold of the pair of mass = 50 meV. Excited 
Xe number density $=10^{18}$cm$^{-3}$.
Two different frequency dependences of the
laser peak intensity, the flat one (in blue) of magnitude $10^6$W mm$^{-2}$
and the one $\propto \omega^4$ (in purple) are compared.
Time duration = 3 ns.
Complications due to mixing and phase factors, and to
other thresholds are all ignored.
}
   \label{rnpe spectrum}
 \end{center} 
\end{figure*}

Experiments can be performed using typically more than three lasers of
different frequencies; more than
two for excitation to the metastable state
and another for the trigger of RNPE.
The trigger frequency $\omega$ (different from $E_{eg}/2$,
hence mistuned for PSR)
is reset each time for measurements at different
photon energies of the RNPE continuous spectrum
of Fig(\ref{rnpe spectrum}).
Hence it is desirable to use frequency tunable lasers
for the trigger.
This way there is no Gaussian tail suppression at
each detected spectral point of RNPE photon, while the background PSR
is suppressed by the Gaussian tail factor due to a mismatch
away from $\omega = E_{eg}/2$.
Moreover, the energy resolution of RNPE photon is
essentially determined by the precision of triggering laser
frequency, and not by detected photon energy resolution.
This is a key for success of the precision neutrino
mass spectroscopy, which must resolve photon energies at 
the $\mu$eV level.

\begin{figure*}[htbp]
 \begin{center}
 \epsfxsize=0.6\textwidth
 \centerline{\epsfbox{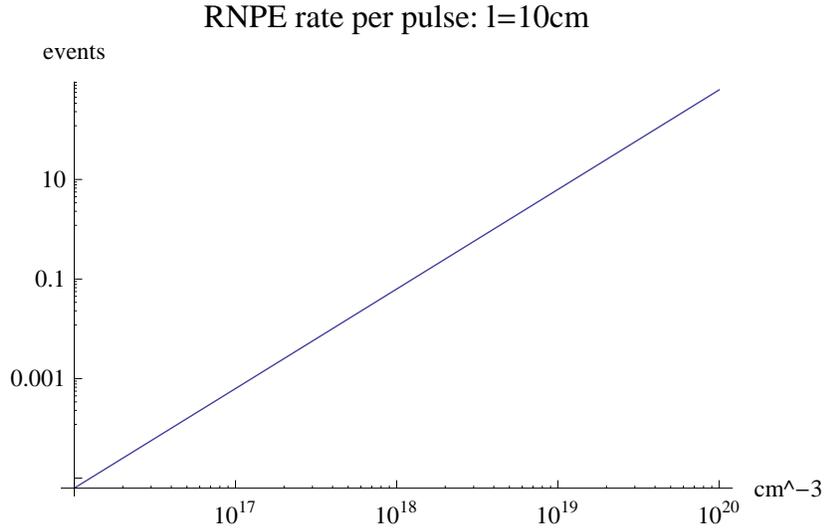}} \hspace*{\fill}
   \caption{RNPE rate (event number per pulse) vs the target number density
of pair emission of m= 1 meV evaluated at
m=50 meV pair threshold. 
Laser intensity
$10^6$W mm$^{-2}$ and time duration 3 ns are assumed.
}
   \label{rnpe rate vs n}
 \end{center} 
\end{figure*}

Theoretical estimates readily  give coherent
RNPE rates as large as, of order $0.1$ events per pulse
for Xe atoms of number density of order $10^{18}$cm$^{-3}$,
which may be realized in solid matrix environments.
With a repetition cycle of $O[1]$ msec interval,
this gives a detectable effective rate of $O[100]$ Hz.
We show the continuous single photon
energy spectrum of Xe macro-coherent RNPE
in Fig(\ref{rnpe spectrum}).
The sharp rise at the threshold is characteristic
of the three-body decay under the momentum conservation,
as is familiar in the $\mu$ decay.
The increasing rate towards the low energy photon,
given in the blue curve of Fig(\ref{rnpe spectrum}),
is due to the flat frequency dependence of the peak intensity
of the trigger laser, and
the decreasing rate in purple is due to $\propto \omega^4$ peak
intensity dependence more akin to the phase space of
the 3-body  spontaneous decay of elementary particles.
The low energy side of the photon energy spectrum
is thus sensitive to the frequency dependence of
the trigger laser intensity, and is inevitably tied
to experimental apparatus used.
The rate scales as $n^2 V$,
with the target number density $n$ and the volume $V$
of coherent region, as shown in Fig(\ref{rnpe rate vs n}).

The problem against the precision neutrino
mass spectroscopy is not the rate itself, if
a sufficient number of target atoms is prepared.
Rather, the serious problem appeared to be in
the signal to the background ratio (S/N) where
the main background source is the physical process of
two photon emission, in particular $N^2$ enhanced PSR.
We discuss this problem shortly.
Higher order QED processes that could sneak into
our photon energy region might appear problematic, but they
are actually negligible if they have (spontaneous) photon emission rates smaller
than the elementary decay rate of metastable (lifetime $> O[1]$ msec in
our standard) state,
since in actual experiments a cycle of measurement for
RNPE is terminated much earlier than the lifetime 
of metastable atom, and the target preparation is recycled.
The single photon SR, entirely outside the photon energy
region of our interest, is negligible also in rate
by the choice of trigger frequency for RNPE.

As shown in the present paper, the S/N problem caused by PSR is much
relaxed and the value of S/N is enormously
enhanced by two effects;
(1) mistuned trigger frequencies for PSR and (2) soliton formation.
We first discuss the first issue.
S/N is increased by the choice of
trigger frequency set at $\omega \leq E_{eg}/2- 2m_1^2/E_{eg}$ of the lightest neutrino pair
emission for RNPE
(we are ordering neutrino masses according to
$m_3 > m_2> m_1$). There is a mismatch of energy from the
largest rate point at $\omega = E_{eg}/2$ of  PSR
by the amount $2m_1^2/E_{eg}$.
If one may assume the Gaussian frequency distribution
for the trigger laser, the
PSR background rate is suppressed by a large factor of
$\exp[- 4 m_1^4/(E_{eg}^2 \delta^2)]$, with
$\delta$ the Gaussian frequency width.
A numerical factor of practical importance is
$ \sqrt{E_{eg} \delta}/2 \sim 1.2 {\rm meV}\sqrt{\delta/{\rm GHz}}$.
It is thus important to have the frequency width of sub-GHz 
in order to explore the sub-meV neutrino mass range.

In Fig(\ref{rnpe rate vs n-mass}) the spectral
RNPE rate of assumed neutrino mass threshold
is plotted, along with the background PSR rate,
taking for the input laser 1 GHz frequency width of Gaussian distribution.
It is estimated from this and similar figures
of the pair mass range, $m_1 = (0.5 \sim 1)\,, m_2 = (8 \sim 10)$ meV,
that this quality of laser can explore the neutrino mass
of order 1 meV.
With an even narrower width laser,
one may hope to explore the sub-meV region and
detection of the relic neutrino \cite{my-taka}.
This, however, rests with the Gaussian nature
of the tail part of the laser frequency distribution,
which must be carefully examined from
experimental points.

Soliton formation gives 
another great merit in which the background
PSR is blocked, completely in ideal situations, 
and may open a new path towards
a controlled RNPE experiment.
In practice, only a partial blocking can be expected due to
various kinds of relaxation processes.
Naively, one would expect that the ratio
of the relaxation to the soliton formation rate
is a relevant quantity for
the partial blocking.
The soliton formation rate is however difficult to estimate
without detailed experimental design and experimental R \& D works.
Study of decoherence in solid environments
by means of detailed numerical simulation and 
R \& D works
is also very important for realization
of the neutrino mass spectroscopy.

\vspace{1cm}
{\bf Acknowledgements}

I should like to thank N. Sasao and
members of SPAN experimental group for informative and helpful
discussions on this subject.

%%%%%%%%%%%%%%%%%%%%%%%%%%%%%%%%%%%%%%%%%%%%%%%%%%%

\vspace{1cm}

{\bf \lromn8 Appendix \lromn1 \; Derivation of effective two level model
}

%\vspace{0.5cm}
\begin{itemize}
\item 
{\bf Atomic system}

The state vector of an atom can be expanded in terms of the wave function,
\begin{eqnarray}
&&
| \psi(t) \rangle = \sum_j c_j(t) e^{-i E_j t}|j \rangle
+ c_g(t) e^{-i E_g t}|g \rangle + c_e(t) e^{-i E_e t}|e \rangle
\,.
\end{eqnarray}
$ c_{\alpha}(t)$ are probability amplitudes in an interaction picture.

The atomic system may interact with fields.
The electric field $\epsilon(x,t)$ that appears in 
the Hamiltonian via E1 or M1 transition
is assumed to have one component alone, namely 
we ignore effects of field polarization.
This is a valid approach under a number of circumstances.
One then decomposes the real field variable $\epsilon(x,t)$ into 
Fourier series, $e^{i\omega t}$ times a complex envelope amplitude $E_0(x,t)$,
and its conjugate, where $E_0(x,t) e^{ikx}$ ($k =$ wave number) is assumed slowly varying in time,
\begin{eqnarray}
&&
\epsilon(x,t) = E_0(x,t)e^{i\omega t} + E_0^*(x,t)e^{-i\omega t}
\,.
\end{eqnarray}

The Schr\"{o}dinger equation for a single atom
\begin{eqnarray}
&&
i \frac{\partial }{\partial  t}|\psi(t) \rangle = (H_0 +
d \epsilon )|\psi(t) \rangle
\,,
\end{eqnarray}
gives the upper level amplitude $c_j(t)$.
Using
\begin{eqnarray}
&&
i \frac{\partial }{\partial  t} \langle j  |\psi(t) \rangle
= \langle j |(H_0 + d \epsilon )|\psi(t) \rangle
\,,
\end{eqnarray}
one has
\begin{eqnarray}
&&
i \frac{dc_j}{dt}e^{- iE_j t}
=
(d_{je} c_e e^{- iE_e t}+d_{jg} c_g e^{- iE_g t}) \epsilon
\,.
\end{eqnarray}
This can formally be
integrated to
\begin{eqnarray}
&&
\hspace*{-1cm}
c_j(t) = -i \int_0^t dt' \left(
d_{je}\epsilon(x,t') c_e(t') e^{i (E_j - E_e)t'}
+ d_{jg}\epsilon(x,t') c_g(t') e^{i (E_j - E_g)t'}
\right)
\nonumber \\ &&
= - i \int_0^t dt' \left( d_{je} c_e(t') e^{i E_{je}t'} + d_{jg} c_g(t') e^{i E_{jg}t'}
\right)
\left( E_0(x,t')e^{i\omega t'} + E_0^*(x,t')e^{-i\omega t'}\right)
\,,
%\nonumber \\ &&
\label{upper level coeff}
\end{eqnarray}
(the initial condition $c_j(0) = 0$ is assumed).

We extend the system of a single atom to a collection of atoms, regarding
variables $c_{\alpha}(x, t)$ as
functions of two variables, $x$ and $ t$.
The population and the coherence of the entire atomic system is described
by the density matrix elements,
 $R_{\alpha \beta}(x,t)\,, \alpha \beta= e, g, j$, which is
the squared product of  wave functions (bilinears
in $c_e\,, c_g\,, c_j$ and their conjugates) and the target number
density $n(x)$.
The macro-variables are defined by
\begin{eqnarray}
&&
R_{\alpha \beta}(x, t) = n(x) c_{\alpha}^*(x,t)c_{\beta}(x,t) %e^{-iE_{\beta \alpha}t}
\,.
\end{eqnarray}
For a notational simplicity 
we often omit space coordinate ($x$)
dependence of probability amplitudes and write these 
simply as $c_{\alpha}(t)$ instead of more proper $c_{\alpha}(x, t)$.

\item 
{\bf Markovian approximation and SVEA} 

The basic strategy of deriving equations for the lower two level amplitudes $c_e, c_g$
in a closed form is 
to eliminate atomic variables related to the upper level amplitude $c_j$.
This is essentially done by neglecting a long-time
memory effect (the Markovian approximation) and making slowly varying envelope
approximation (SVEA) in the terminology of \cite{narducci}.
The idea 
is to replace dynamical variables, $c_e(t'), c_g(t'), E(x,t')$ in the integrand
of eq.(\ref{upper level coeff}), by their 
values at time $t$, neglecting all the past memory effects.
This gives 
\begin{eqnarray}
&&
%\hspace*{-1cm}
c_j(t) \approx d_{je}c_e \left( 
\frac{1 - e^{i(\omega + E_{je})t}}{\omega + E_{je}}E_0 
-  \frac{1 - e^{- i(\omega - E_{je})t}}{\omega - E_{je}}E_0^*\right)
\nonumber \\ &&
+ d_{jg}c_g \left( \frac{1 - e^{i(\omega + E_{jg})t}}{\omega + E_{jg}}E_0 
-  \frac{1 - e^{- i(\omega - E_{jg})t}}{\omega - E_{jg}}E_0^*\right)
\,,
%\nonumber \\ &&
\label{upper level amp}
\end{eqnarray}
which is inserted into
equations for the lower levels 
\begin{eqnarray}
&&
\frac{dc_e}{dt}
= - i \sum_j d_{ej} \epsilon(x,t)c_j(t) e^{i E_{ej}t}
\,,
\\ &&
\frac{dc_g}{dt}
= - i \sum_j d_{gj} \epsilon(x,t)c_j(t) e^{i E_{gj} t}
\,.
\end{eqnarray}

Unitarity, namely the probability conservation, given by
\begin{eqnarray}
&&
\frac{d}{dt} (|c_e|^2 + |c_g|^2 + \sum_j |c_j|^2 ) = 0
\,,
\end{eqnarray}
does not hold when limited to two amplitudes $c_e\,, c_g$ alone,
namely in the Markovian approximation.
However there exists an effective conservation law that
holds for the two level system, given later by (\ref{effective conservation}).
There seems some misunderstanding on this point in \cite{narducci}.

We further neglect rapidly oscillating terms assuming the
nearly, but not necessarily exactly, tuned condition, $\omega \approx E_{eg}/2$, which amounts to
\begin{eqnarray}
&&
\frac{dc_e}{dt}
 = i \left( \mu_{ee}|E_0^2|c_e(t) +  \mu_{eg}( E_0^*\,)^2 c_g(t)e^{-i (2\omega - E_{eg})t}
\right)
\,,
\\ &&
\frac{dc_g}{dt}
= i \left(  \mu_{gg}|E_0^2| c_g(t) +  \mu_{ge}E_0^2 c_e(t) e^{i (2\omega - E_{eg})t}
\right)
\,,
\\ &&
 \mu_{ee} = 2 \sum_j \frac{d_{je}^2E_{je}}{E_{je}^2 - \omega^2}
\,, \hspace{0.5cm}
 \mu_{gg} = 2 \sum_j \frac{d_{jg}^2E_{jg}}{E_{jg}^2 - \omega^2}
 \,,
\\ &&
 \mu_{eg} = \sum_j \frac{d_{je}d_{jg} }{E_c - \delta \omega}
\,, \hspace{0.5cm}
 \mu_{ge} = \sum_j \frac{d_{je}d_{jg} }{E_c + \delta \omega}
 \,,
\end{eqnarray}
where $E_c = \frac{1}{2}(E_{jg} + E_{je})
\,, \;
\delta \omega = \omega - \frac{E_{eg}}{2} $ as in the text.
All of $d_{j\alpha}\,,  \mu_{\alpha \beta}$ are taken as real.

A mistake in \cite{narducci}
is that $ \mu_{eg} =  \mu_{ge}$ is assumed even at
$\delta \omega \neq 0$. This equality holds
only at the middle point of frequency
$\omega = E_{eg}/2$.
Thus, results of \cite{narducci} away from this
tuned frequency should be taken with skepticism.

Writing this equation in a matrix form, 
\begin{eqnarray}
&&
\frac{d}{dt} \left(
\begin{array}{c}
c_{e}  \\
c_{g}  
\end{array}
\right)= - i{\cal H}\left(
\begin{array}{c}
c_{e}  \\
c_{g}  
\end{array}
\right)
\,,
\\ &&
{\cal H} =
\left(
\begin{array}{cc}
 \mu_{ee}|E_0^2| & e^{-i (2\omega - E_{eg})t} \mu_{eg}( E_0^*\,)^2  \\
e^{i (2\omega - E_{eg})t} \mu_{ge}E_0^2 &   \mu_{gg}|E_0^2|
\end{array}
\right)
\,,
\end{eqnarray}
one finds that the effective
Hamiltonian ${\cal H}$ becomes hermitian only by neglecting
$\delta \omega/E_c$ terms in $ \mu_{eg}\,, \mu_{ge}$.

We introduce symmetric and anti-symmetric functions of
$\delta \omega = \omega - E_{eg}/2$ as
\begin{eqnarray}
&&
\mu^+ = \sum_j \frac{d_{je}d_{jg} E_c}{E_c^2 - \delta \omega^2}
\,, \hspace{0.5cm}
\mu^- = \sum_j \frac{d_{je}d_{jg} \delta \omega}{E_c^2 - \delta \omega^2}
\,,
\\ &&
 \mu_{eg} = \mu^+ + \mu^-
\,, \hspace{0.5cm}
 \mu_{ge} = \mu^+ - \mu^-
 \,,
\end{eqnarray}
and write
\begin{eqnarray}
&&
{\cal H} = {\cal H}^+ + {\cal H}^-
\,,
\\ &&
{\cal H}^+ =
\left(
\begin{array}{cc}
 \mu_{ee}|E_0^2| & e^{-i (2\omega - E_{eg})t}\mu^+( E_0^*\,)^2  \\
e^{i (2\omega - E_{eg})t}\mu^+E_0^2 &   \mu_{gg}|E_0^2|
\end{array}
\right)
\,,
\\ &&
{\cal H}^- =
\left(
\begin{array}{cc}
0 & e^{-i (2\omega - E_{eg})t}\mu^-( E_0^*\,)^2  \\
- e^{i (2\omega - E_{eg})t}\mu^- E_0^2 &  0
\end{array}
\right)
\,.
\end{eqnarray}
${\cal H}^+$ is hermitian, while ${\cal H}^-$ is anti-hermitian.
The anti-symmetric piece ${\cal H}^-$ vanishes at the middle point
of $\delta \omega = 0$, since
${\cal H}^- \propto \mu^- \propto \delta \omega$,
hence is small except where $\delta \omega$ is large of $  O[E_c]$.
We shall ignore effect of ${\cal H}^-$ and assume $\mu_{ge} = \mu_{eg}$
in the main text of the present work.

\vspace{0.5cm}
\item
{\bf Generalized Bloch vector and its dynamical equation} 

The  4 component Bloch vector is defined by
\begin{eqnarray}
&&
R_0 = n (|c_e|^2 + |c_g|^2)
\,,
\\ &&
R_1 = i n (c_g^*c_e e^{i\eta } - c_e^* c_g e^{-i\eta })
\,,
\\ &&
R_2 = - n (c_g^*c_e e^{i\eta } + c_e^* c_g e^{-i\eta })
\,,
\\ &&
R_3 = n (|c_e|^2 - |c_g|^2)
\,,
\\ &&
\eta = (2\omega -E_{eg}) t - 2kx + 2\varphi
\,,
\label{field phase}
\end{eqnarray}
where we assume a standard form of field,
\begin{eqnarray}
&&
\epsilon(x, t) = \epsilon_0(x, t) \cos (\omega t - kx + \varphi)
\,,
\end{eqnarray}
with  $k = \pm \omega$.
The real amplitude $\epsilon_0(x, t)$ and
the phase $ \varphi(x, t)$ are assumed both slowly varying in time and
in space.
Note the relation of real and complex field,
$\epsilon_0^2 = 4|E_0|^2$ ($E_0$ is complex including $e^{i(\omega t - kx)}$).

The generalized Bloch equation is given by
\begin{eqnarray}
&&
\hspace*{-2cm}
\frac{\partial}{\partial t}R_1 
= \left( \frac{ \mu_{ee} -  \mu_{gg}}{4}\epsilon_0^2 + (2\omega - E_{eg} + 2 \frac{\partial \varphi}{\partial t})
\right)R_2 + \frac{\mu^+}{2} \epsilon_0^2R_3 - \frac{\mu^-}{2} \epsilon_0^2 R_0
\,,
\\ &&
\frac{\partial}{\partial t}R_2 = - \left( \frac{ \mu_{ee} -  \mu_{gg}}{4}\epsilon_0^2 + (2\omega - E_{eg} + 2 \frac{\partial \varphi}{\partial t})
\right)R_1
\,,
\\ &&
\frac{\partial}{\partial t}R_3 = - \frac{\mu^+}{2} \epsilon_0^2 R_1
\,,
\\ &&
\frac{\partial}{\partial t}R_0 
= \frac{\mu^-}{2} \epsilon_0^2 R_1
\,,
\\ &&
\mu^+ = \sum_j \frac{d_{ej}d_{gj}E_c}{E_c^2 - (\omega - E_{eg}/2)^2}
\,,
\end{eqnarray}
proved by using the Schr\"{o}dinger equation for $c_e, c_g$.
Note that in RHS of these equations there is no phase factor like
$e^{i(2kt - 2 \omega t + E_{eg}t)}$.

The conservation law is extended to the 4-vector;
\begin{eqnarray}
&&
\frac{\partial}{\partial t} (  R_0^2 + R_1^2 + R_2^2 + R_3^2) = 0
\,.
\label{effective conservation}
\end{eqnarray}
Only at the middle point of $\omega = E_{eg}/2$
this conservation reduces to the usual type of
conservation for the 3-vector norm,
$\frac{\partial}{\partial t} ( R_1^2 + R_2^2 + R_3^2) = 0$
(the assumption taken in \cite{narducci}),
since in this case $\mu^- = 0$ and one has separately
$\partial_t R_0^2 = 0$.

\vspace{0.5cm}
\item
{\bf Polarization of medium}

Polarization vector is defined by
\begin{eqnarray}
&&
P = n(x) \langle \psi(t) | d | \psi(t)  \rangle 
%\nonumber \\ && \hspace*{-2cm}
=
n \sum_j ( d_{ej} c_e^* c_j e^{-iE_{je}t} + d_{gj} c_g^* c_j e^{-iE_{jg}t} + {\rm c.c})
\,,
\end{eqnarray}
for which we eliminate $ c_j $ using eq.(\ref{upper level amp}).

The polarization can be decomposed into the in-phase $\cos(\omega t - kx + \varphi)$ and 
the out-phase $\sin (\omega t - kx + \varphi)$ parts;
\begin{eqnarray}
&&
P = n(x) ( \mu_{ee}|c_e|^2 +  \mu_{gg}|c_g|^2 -  ( \mu_{eg}c_g c_e^* e^{-i\eta} +  \mu_{ge}c_g^* c_e e^{i\eta})) 
\epsilon_0(x, t)\cos(\omega t - kx + \varphi)
\nonumber \\ &&
- i ( \mu_{eg}c_g c_e^* e^{-i\eta} -  \mu_{ge}c_g^* c_e e^{i\eta}))\epsilon_0(x, t)\sin(\omega t - kx + \varphi)
\nonumber  \\ &&
\hspace*{-1cm}
= (\frac{ \mu_{ee} +  \mu_{gg}}{2}n + \frac{ \mu_{ee} -  \mu_{gg}}{2}R_3 + \mu^+ R_2)
\epsilon_0(x, t)\cos(\omega t - kx + \varphi) 
%\nonumber  \\ &&
+ \mu^+ R_1\epsilon_0(x, t)\sin(\omega t - kx + \varphi) 
\,.
\nonumber  \\ &&
\end{eqnarray}
$O[\mu^- ]$ terms do not contribute to hermitian polarization $P$.

The  Maxwell equation
\begin{eqnarray}
&&
(\frac{\partial^2}{\partial x^2} - \frac{\partial^2}{\partial t^2})\epsilon
= \frac{\partial^2 P}{\partial t^2}
\,,
\end{eqnarray}
gives for envelope amplitude and phase variation under SVEA
\begin{eqnarray}
&&
(\partial_t + \partial_x)\epsilon_0^2 = \omega \mu^+ \epsilon_0^2 R_1
\,,
\\ &&
\hspace*{-1cm}
(\partial_t + \partial_x) 2\varphi = \omega
\left( - \omega \mu^+ R_2 + \frac{ \mu_{ee} -  \mu_{gg}}{2}R_3 + 
\frac{ \mu_{ee} +  \mu_{gg}}{2}n  \right) 
\,.
\end{eqnarray}

In order to simplify equations, we introduce new variables by
\begin{eqnarray}
&&
R_2'  = \frac{R_2 - \gamma R_3}{\sqrt{1 + \gamma^2}} 
\,, \hspace{0.5cm}
R_3'  = \frac{\gamma R_2 + R_3}{\sqrt{1 + \gamma^2}} 
\,,
\\ &&
\omega_R = \frac{\sqrt{1 + \gamma^2}} {2} \mu^+ \epsilon_0^2 
\,,
\\ &&
\gamma = \frac{ \mu_{ee} -  \mu_{gg}}{2\mu^+}
\,, \hspace{0.5cm}
\Omega = 2 \omega - E_{eg} + 2 \frac{\partial \varphi}{\partial t} 
\,.
\end{eqnarray}

The basic set of equations is given by
\begin{eqnarray}
&&
\partial_t R_1 = \omega_R R_3' 
+ \frac{\Omega}{\sqrt{1 + \gamma^2}} (R_2' + \gamma R_3')
- \frac{\mu^-}{2} \epsilon_0^2 R_0
\,,
\\ &&
\partial_t R_2' = - \frac{\Omega}{\sqrt{1 + \gamma^2}}R_1
\,,
\\ &&
\partial_t R_3' = - \omega_R R_1 - \frac{ \gamma\Omega}{\sqrt{1 + \gamma^2}}R_1
\,,
\\ &&
\partial_t R_0 = \frac{\mu^-}{2} \epsilon_0^2 R_1
\,,
\\ &&
(\partial_t + \partial_x) \omega_R + \kappa \omega_R = \omega \mu^+ \omega_R R_1 
\,,
\\ &&
(\partial_t + \partial_x) \Omega = 2\omega \mu^+ R_1 \Omega 
\,,
\end{eqnarray}
from which it follows
\begin{eqnarray}
&&
\partial_t \left(  R_0^2 + R_1^2 + (R_2')^2 + (R_3')^2
\right) = 0
\,.
\end{eqnarray}

We have not included atomic relaxation effects given by
parameters, $T_1\,, T_2\,, T_2^*$ \cite{sr review}.

One can consistently take $\Omega = 0$, namely
$ \partial_t \varphi = E_{eg}/2 - \omega$.
This adjustment of field phase is assumed in the text of this paper.

For both $\mu^- = 0$ and $\Omega = 0$
the system further simplifies to
eqs.(\ref{bloch 1}), (\ref{bloch 2}) and (\ref{area relation}) in the text by an appropriate choice of variables.

\end{itemize}

\vspace{1cm}

{\bf \lromn9 Appendix \lromn2 \; Details towards construction of analytic solutions
}

We start from discussions that lead to
introduction of the tipping angle $\theta (x,t)$, eq.(\ref{tipping angle def}), 
related to
the Bloch vector component by $R_3 \propto \cos \theta$.
Another important relation (\ref{area relation}), $\partial_t \theta \propto |E_0^2|$, 
gives
a physical content of the area function $\theta$,  relating it to an integral of
the pulse flux $|E_0^2|$.

Our method for solving non-linear partial differential equations 
of two independent variables
($x$ and $t$)
is to integrate in one variable $t$ and replace integration constants 
obtained this way by functions including
another variable $x$.
The method  works for our problem of one mode propagation,
but it is not a general mathematical method.

At finite time $t$ we allow the integration constant $t_0,\theta_0$ of 0d solution
(solution without space dependence), eq.(\ref{area-no-space}),
to vary in spacetime according to 
\begin{eqnarray}
&&
t_0 \rightarrow T(t-x)
\,, \hspace{0.5cm}
\theta_0 \rightarrow \theta_p(x)
\,,
\\ &&
( \partial_t + \partial_x)\theta  + \alpha (\cos \theta - \cos \theta_p ) = 0
\,,
\end{eqnarray}
noting a trivial equality,
$(\partial_t + \partial_x)T(t-x) = 0$.
Hence, solutions are written in terms of two functions to be determined by
the initial and the boundary data,
\begin{eqnarray}
&&
\hspace*{-2cm}
\theta(x\,, t) = {\rm arccos}\;
\frac{\cos \theta_p(x)\cosh \left( \alpha (t - T(t-x)\,)\sin \theta_p\right) - 1}
{\cosh \left( \alpha (t - T(t-x)\,)\sin\theta_p\right) - \cos \theta_p(x)}
\,,
\\ &&
\hspace*{-1cm}
\omega_R (x, t) = \partial_t \theta = 
\frac{- \alpha (1 - \partial_t T(t-x)\,)\sin^2 \theta_p(x)}{\cosh \left( \alpha(t - T(t-x)\,) \sin\theta_p\right) - \cos \theta_p(x)}
\,,
\label{pulse strength}
\end{eqnarray}
where $\theta_p(x)\,, T(y=t-x) $ are yet to be determined.

The following, somewhat complicated steps leading to
eqs.(\ref{boundary condition}) and (\ref{initial condition 2}) are processes of how the initial and the boundary
conditions determine the unknown functions, $\theta_p(x)\,, T(y) $. 

The given boundary data at some spatial point
$x=0$, the target end at which laser irradiation takes place, and the initial data
at $t= 0$ are 
\begin{eqnarray}
&&
\theta (x=0, t) \equiv \tilde{\theta}(t)
\,, \hspace{0.5cm}
\theta(x, t=0) 
\,.
\label{input data}
\end{eqnarray}
These are related to the variable $\omega_R (x, t)$  by
using $\partial_t \theta(x,t) = \omega_R(x,t)$,
\begin{eqnarray}
&&
\theta (0, t)= \int_{-\infty}^{t}dt' \omega_R(0,t')
\,,
\\ &&
\theta(x,0) = \int_{-\infty}^0 dt' \omega_R(x,t')
\,.
\end{eqnarray}
Note that two data (\ref{input data}) are independent.

We may solve for the unknown function $T(\tau)$
using the boundary condition, to get
\begin{eqnarray}
&&
\hspace*{-1cm}
T(\tau) = \tau - \frac{1}{\alpha\sin \theta_0} {\rm arccosh}\, \frac{1- \cos \theta_0 \cos \tilde{\theta}(\tau) }
{\cos \theta_0 - \cos \tilde{\theta}(\tau) }
\,, \hspace{0.5cm}
\theta_0 \equiv \theta_p(0) 
\,.
\label{boundary condition}
\end{eqnarray}
The solution for $\theta_p(x)$ is obtained from the initial condition,
given by
\begin{eqnarray}
&&
\partial_t\tilde{\theta} (-x) (\cos \theta_p - \cos \theta(x,0)\,)=  (\cos \theta_0 - \cos\tilde{\theta} (-x)\,)
\omega_R(x,0)
\,.
\label{initial condition}
\end{eqnarray}

This equation together with
eq.(\ref{pulse strength}) calculated at $t=0\,, x=0$ and
(\ref{boundary condition}),
gives, with $\tilde{\mu} \equiv \sqrt{(\mu_{gg}-\mu_{ee})^2 + 4 \mu_{ge}^2}/4$, 
\begin{eqnarray}
&&
\cos \theta (0,0) =  \pm \cos \tilde{\theta}(0)
\,,
\hspace{0.5cm}
\tilde{\theta}(0) = \tilde{\mu}
\int_{-\infty}^{0} dy |\epsilon_0^2(y)|
\,,
\label{initial area}
\end{eqnarray}
+(-) corresponding to amplifier (absorber).
The second equation of (\ref{initial area})
is derived from the area-intensity relation,
$\partial_t \tilde{\theta} \propto |\epsilon_0^2|$, for the incident pulse.

If the major part of input laser is still far away from the target end of $x=L$,
$\tilde{\theta}(0) \approx 0$,
and $\theta(0,0) \approx 0 \,(\pi)$.
The amplifier case of $\theta(0,0) \approx 0$ corresponds to a physical situation in which
medium is excited by other lasers, while the absorber case of $\theta(0,0) \approx \pi$
to medium in the ground state.

Equation (\ref{pulse strength}) calculated at other points of $t=0$
gives 
\begin{eqnarray}
&&
\cos \theta (x,0) = \cos \theta_p(x) 
- \frac{\sin^2 \theta_p(x)}{\cosh(\,\alpha T(-x) \sin\theta_p(x)  \,) 
- \cos \theta_p(x)}
\,.
\label{initial condition 2}
\end{eqnarray}
Since $T(-x)$ here is already given in terms of $\tilde{\theta}(-x)$
by eq.(\ref{boundary condition}),
this equation determines $\cos \theta_p(x)$ in terms of the
initial data $\cos \theta (x,0)$.
We find it possible to construct solutions of these equations only when
$\theta_p(x) = \theta_0$ (spatially constant).

Explicit form of solution is then ($0 \leq \theta_0 \leq \pi$)
\begin{eqnarray}
&&
\hspace*{-1cm}
|E_0^2(x,t)| = \frac{\sin^2 \theta_0 |\epsilon_0^2(t-x)|}
{\cosh (\alpha x\sin \theta_0 ) (1- \cos \theta_0 \cos \tilde{\theta}) 
\pm \sinh (\alpha x\sin \theta_0 ) \sin \theta_0\sin \tilde{\theta} - 
\cos \theta_0 (\cos \theta_0 - \cos \tilde{\theta})}
\,,
\label{general sol 1}
\nonumber 
\\ &&
\\ &&
\hspace*{-1cm}
\cos \theta (x,t) = \mp \frac{\cosh (\alpha x\sin \theta_0 )\cos \theta_0
(1 - \cos \theta_0 \cos \tilde{\theta})
\pm \sinh (\alpha x \sin \theta_0 ) \cos \theta_0 \sin \theta_0\sin \tilde{\theta}
-  (\cos \theta_0 - \cos \tilde{\theta})}
{\cosh (\alpha x \sin \theta_0 ) (1- \cos \theta_0 \cos \tilde{\theta}) 
\pm \sinh (\alpha x \sin \theta_0 ) \sin \theta_0\sin \tilde{\theta} - 
\cos \theta_0 (\cos \theta_0 - \cos \tilde{\theta})}
\,,
\nonumber 
\\ &&
\label{general sol 3}
\\ &&
\hspace*{1cm}
R_3 (x,t) = \frac{n}{\sqrt{1+ \gamma^2}} \cos \theta (x,t)
\,,
\end{eqnarray}
with variable dependence given by $x$ explicitly and $t-x$ in
$\tilde{\theta}(t-x) $.
These solutions are given in terms of the strength of input pulse,
\begin{eqnarray}
&&
\tilde{\theta}(t-x) =
\tilde{\mu}
\int_{-\infty}^{t-x} dy |\epsilon_0^2(y)|
\,.
\end{eqnarray}
The constraint $\cos^2 \theta (x,t) \leq 1$ is satisfied for any $\theta_0$.

Note that $\tilde{\theta}$ is monotonically increasing function of its argument.
The solution  $\theta(x,t)$ is valid only for $\partial_t \theta > 0$
for the positivity of the pulse strength.

We do not know whether these solutions have complete generality
and no other solutions exist, but
they seem to be adequately general for our purposes.
The initial data $\theta (x,0)$ at $x>0$ consistent with these solutions
is given in terms of the input area at negative arguments $\tilde{\theta}(-x)$,
and they are independent of the boundary data $\theta (0,t)$ 
at $t> 0$ given by the same area of
positive arguments.
One might consider the situation of $\tilde{\theta}(-x) = 0 $ at $x>0$, which
gives
\begin{eqnarray}
&&
\cos \theta (x,0)
= 
\frac{\cos \theta_0\cosh (\alpha x\sin \theta_0 ) +1}
{\cosh (\alpha x \sin \theta_0 ) +
\cos \theta_0 }\,,
\end{eqnarray}


\vspace{1cm}
\begin{thebibliography}{99}

\bibitem{sr review}
For an excellent review of both the theory and
experiments of superradiance,
M. Benedict, A.M. Ermolaev, V.A. Malyshev, I.V. Sokolov, and
E.D. Trifonov,
{\it Super-radiance Multiatomic coherent emission},
Informa (1996).

For a formal aspect of the theory,
M. Gross and S. Haroche, {\it Phys.Rep.}{\bf 93}, 301(1982).

The original suggestion of superradiance is due to
R.H. Dicke, {\it Phys. Rev.}{\bf 93}, 99(1954).

\bibitem{macro-coherence}
M. Yoshimura, C. Ohae, A. Fukumi, K. Nakajima, I. Nakano,
H. Nanjo, and N. Sasao,
{\it Macro-coherent two photon and radiative
neutrino pair emission}, arXiv 805.1970[hep-ph](2008).

M. Yoshimura, {\it Neutrino Spectroscopy using Atoms (SPAN)},
in Proceedings of 4th NO-VE International Workshop,
edited by M. Baldo Ceolin(2008).




\bibitem{narducci} L.M. Narducci et al.,
{\it A Model of Degenerate Two-photon Amplifier},
in {\it Cooperative Effects in Matter and Radiation}, ed. by C.M. Bowden, D.W. Howgate, and H.R. Robl
Prenum Press, New York (1977);
L.M. Narducci, W,W. Eidson, P. Furcinitti, and D.C. Eteson,
{\it Phys. Rev.}{\bf A 16}, 1665 (1977).



\bibitem{my-06}
M. Yoshimura,
{\it Phys. Rev.}{\bf D75}, 113007(2007).


\bibitem{pv}
M. Yoshimura, A. Fukumi, N. Sasao and T. Yamaguchi,
{\it Progr. Theor. Phys.}{\bf 123}, 523(2010),
and {\it Parity violating observables
in radiative neutrino pair emission from metastable atoms},
arXiv:0907.0519v2 [hep-ph] (2009).

\bibitem{coherent light propagation in 2 level}
S.L. McCall and E.L. Hahn,
{\it Phys. Rev.}{\bf 183}, 457(1969).

For a review,
L. Allen and J.H. Eberly,
{\it Optical Resonance and Two-level Atoms},
Dover, New York, (1975).

For comparison with experimental results,
R.E. Slusher and H.M. Gibbs,
{\it Phys. Rev.}{\bf A4}, 1634(1972).


\bibitem{doppler broadening}
For examples,
B.H. Bransden and C.J. Joachain,
{\it Physics of Atoms and Molecules},
2nd edition, Prentice Hall (2003);
D. Budker, D.F. Kimball and D.P. DeMille,
{\it Atomic Physics}, Oxford University Press, New York
(2004).




\bibitem{dark state}
For a review,
C. Cohen-Tannoudji, J. Dupont-Roc, and G. Grynberg,
{\sl Atom-Photon Interactions}, Wiley-VCH(2004).


\bibitem{my-taka}
T. Takahashi and M. Yoshimura,
{\it Effect of Relic Neutrino on Neutrino Pair Emission
from Metastable Atoms},
hep-ph/0703019.

%\bibitem{2 pulse potential}

%\bibitem{}

\end{thebibliography}
\end{document}